\documentclass[%
aip,
amsmath,amssymb,
reprint,%
]{revtex4-1}

\usepackage{graphicx}% Include figure files
\usepackage{dcolumn}% Align table columns on decimal point
\usepackage{bm}% bold math
\usepackage[utf8]{inputenc}
\usepackage[T1]{fontenc}
\usepackage{mathptmx}
\usepackage{amsfonts}
\usepackage{etoolbox}
\usepackage{xcolor}
\usepackage{lineno}
\usepackage{amsfonts,amsmath,amssymb,amsthm,relsize}
\usepackage{hyperref}
\def\beq{\begin{equation}}
\def\eeq{\end{equation}}
\def\bea{\begin{eqnarray}}
\def\eea{\end{eqnarray}}
\makeatletter
\def\@email#1#2{%
	\endgroup
	\patchcmd{\titleblock@produce}
	{\frontmatter@RRAPformat}
	{\frontmatter@RRAPformat{\produce@RRAP{*#1\href{mailto:#2}{#2}}}\frontmatter@RRAPformat}
	{}{}
}%
\makeatother

\begin{document}
	
	\preprint{AIP/123-QED}

\title{Resetting mediated navigation of active Brownian searcher in a homogeneous topography}
	% Force line breaks with \\
%	
	\author{Gourab K. Sar}
%	\thanks{mr.gksar@gmail.com}
%	\thanks{Equal contribution}
	\affiliation{%
		Physics and Applied Mathematics Unit, Indian Statistical Institute, 203 B. T. Road, Kolkata 700108, India}

		\author{Arnob Ray}
%	
%\thanks{Equal contribution}
%\thanks{Corresponding author}
%\thanks{arnobray93@gmail.com }
	\affiliation{%
		Physics and Applied Mathematics Unit, Indian Statistical Institute, 203 B. T. Road, Kolkata 700108, India}

	\author{Dibakar Ghosh}
%	\email{dibakar@isical.ac.in}
	\affiliation{Physics and Applied Mathematics Unit, Indian Statistical Institute, 203 B. T. Road, Kolkata 700108, India}		

		\author{Chittaranjan Hens}
%	\thanks{chittaranjanhens@gmail.com}
\affiliation{Center for Computational Natural Sciences and Bioinformatics,
International Institute of Information Technology, Gachibowli, Hyderabad 500 032, India}

	\author{Arnab Pal}
	\thanks{Corresponding author}
	\thanks{arnabpal@imsc.res.in}
\affiliation{The Institute of Mathematical Sciences, IV Cross Road, CIT Campus, Taramani, Chennai 600 113, Tamil Nadu, India.}
\affiliation{Homi Bhabha National Institute, Training School Complex, Anushakti Nagar, Mumbai 400094, India.}
	
	\date{\today}% It is always \today, today,
	
\begin{abstract}
Designing navigation strategies for search time optimization remains of interest in various interdisciplinary branches in science. In here, we focus on microscopic self-propelled searchers namely active Brownian walkers in noisy and confined environment which are mediated by one such autonomous strategy namely resetting. As such, resetting stops the motion and compels the walkers to restart from the initial configuration intermittently according to an external timer that do not require control by the walkers. In particular, the resetting coordinates are either quenched (fixed) or  annealed (fluctuating) over the entire topography. Although the strategy relies upon simple rules, it shows a significant ramification on the search time statistics in contrast to the original search. We show that the resetting driven protocols mitigate the performance of these active searchers based, robustly, on the inherent search time fluctuations. Notably, for the annealed condition, resetting is always found to expedite the search process. These features, as well as their applicability to more general optimization problems starting from queuing systems, computer science to living systems, make resetting based strategies universally promising.
\end{abstract}

\maketitle

%\linenumbers
\section{\label{sec:level1}Introduction}
Active particles operate far from thermodynamic equilibrium. These microswimmers propel themselves with directed motion and thus drive themselves out of equilibrium \cite{toner2005hydrodynamics,SR-review,romanczuk2012active}. In other words, a constant consumption and dissipation of energy results in non-equilibrium activity --   broken detailed balance being a hallmark property of such motion \cite{gnesotto2018broken}. Several living entities such as \textit{E. coli} bacteria are known to perform such active motion \cite{berg2008coli,berg2018random}. In addition, engineered materials such as the Janus particles also work as active autonomous agents to locate and deliver microscopic cargos \cite{jiang2010active}. Moreover, active systems can also provide cues to form spatial organizations, self assemblies through their non-linear interaction, collective motion \cite{martin2018collective}, structural organization and pattern formation \cite{SR-review,gompper20202020,martinez2021active,mijalkov2013sorting,bechinger2016active,volpe2011microswimmers}. No wonder why active systems have become a focal point of recent interests in physics, chemistry and biology.

There has been quite a growing interest in active search processes which have important technological and medical applications, such as cargo transport \cite{ross2008cargo}, targeted drug delivery \cite{ghosh2020active} or identification of infectious protein along a long DNA at the microscopic level \cite{benichou2011intermittent}. Other
potential applications of active search processes span in  biology, computer science, ecology, soft matter and many other fields \cite{martinez2021active,bechinger2016active,wang2017spatial,malakar2018steady,angelani2014first}. With the theoretical and experimental advances side by side on engineering microswimmers, it is only natural to ask how efficient these microscopic devices are to deliver molecular cargo at the desired locations. However, there is a cost in such designing which we assign to be the mean search time to a given target(s) starting from some specific location. Hence, the problem boils down to finding the most efficient route towards a desired target resulting in the lowest search time. Nonetheless, there is no generic pattern on adapting a universal search strategy for all, but drawing insights from various class of search processes (from micro to macro), maybe a class of optimal strategies could be idealized.

Over the years, many realistic models have been proposed to describe and analyze how a searcher spans through the search space looking for targets \cite{benichou2011intermittent,condamin2007first,benichou2005optimal,benichou2010geometry,bray2013persistence}. 
These include Brownian motion, and long range searches such as L\'evy flights and L\'evy walks. In
particular, L\'evy statistics, among others models, have been successfully used to describe the emergence of optimal search strategies in natural systems at different length scales, from molecular entities to foraging animals to human motion patterns \cite{viswanathan2011physics,rhee2011levy,zaburdaev2015levy}. The active search can well be described within the L\'evy strategies since they combine random jumps when the searchers are non-reactive followed by local diffusive jumps when the searchers explore for targets \cite{volpe2017topography}. This is also called intermittent search strategy \cite{benichou2011intermittent}. One notable result in this regard is that the most efficient strategy is a Brownian search (which is a limiting case) while L\'evy strategy seems to be optimal for sparse targets when resources are plentiful \cite{lomholt2008levy,palyulin2014levy}. 

A similar kind of intermittent strategy namely \textit{resetting} has been found to be extremely efficient in optimizing search in many complex processes \cite{evans2011diffusion,evans2011diffusion-opt,evans2020stochastic,gupta2022stochastic,kusmierz2014first,reuveni2016optimal,pal2017first,pal2019first,de2020optimization,pal2016diffusion-K,chechkin2018random,pal2019first-V,bhat2016stochastic,pal2019landau,bressloff2020directed,tal2020experimental,montanari2002optimizing,luby1993optimal,reuveni2014role,ray2020space,ray2021mitigating,roldan2016stochastic,bonomo2021mitigating}. As such, the searcher looks for the targets utilizing its own active motion and if successful, the search is complete. However, the motion can often be interrupted (i.e., stopped at random times) and the searcher is expected to return to its initial location and restart/renew the motion. Thus resetting mixes between the exploration and return phases. Such motion is also prevalent in nature since foraging animals return to their den to rest, bees return to their hive and drones return to their docking station to recharge or refuel \cite{pal2020search}. Considerable progress has been made to understand the effect of variety of ways in which resetting mechanisms and underlying dynamics combine \cite{campos2015phase,kusmierz2015optimal,singh2020resetting,pal2015diffusion,ray2019peclet,huang2021random,gonzalez2021diffusive}. To this end, the universal framework of first passage under restart and its applicability to diffusive processes under resetting is noteworthy \cite{reuveni2016optimal,pal2017first,pal2019first,bonomo2021first}.

In this paper,
our central aim is to understand the effects of resetting based protocols in active search processes. Furthermore, we assume that the active micro-swimmers conduct their search in a confining topography in the presence of target. We explore the possibility of designing different resetting protocols -- one where the searcher is compelled to return to a fixed location (quenched condition) and the other one where the searcher is reset to a random location (annealed condition). As will be shown here, both the geometry of the arena and choice of the resetting protocols play a crucial role in determining the searcher’s motion and in effect to the search efficiency. Our study reveals that resetting can indeed reduce the search time compared to that of the original search process. Secondly, we observe that resetting mechanism works in advantage \textit{when} the fluctuations in the search time for the underlying active process is large -- a robust feature that is observed quite ubiquitously (see also \cite{pal2022inspection}). We distill this core idea  for the randomized/annealed initial conditions where resetting driven search always works efficiently.

We start by looking into the navigation of active Brownian particles (ABP) or micro-swimmers in a two dimensional finite domain in the presence of target (Sec.\ \ref{sec:level2}). We first analyze the search properties of this minimal model in Sec.\ \ref{sec:level3} --- fluctuations and densities of the first passage time are computed. Next, we introduce the autonomous resetting strategies and study the motion of ABPs in Sec.\ \ref{sec:level4}. In particular, we design two different resetting protocols namely annealed and quenched that correlate with the active motion. Using extensive Brownian dynamics simulations, we demonstrate how the new search protocols enhance performances compared to that obtained from the original search process. Lastly in Sec.\ \ref{sec:level5}, we provide analytical support and underpin the core principle that lies beneath the robustness of resetting-mediated strategies. We conclude with future perspectives on this useful intermittent search strategy.

\section{\label{sec:level2}Model Description}

The motion of an ABP in a two-dimensional topography can be modeled by stochastic Langevin equations namely \cite{romanczuk2012active,volpe2014simulation,basu2018active}
\begin{align}
\dot{x}(t) &= v_{0}\cos(\theta (t)), \label{eq.1}\\
\dot{y}(t) &= v_{0}\sin(\theta (t)), \label{eq.2}\\
\dot{\theta}(t) &= \sqrt{2D_{R}} \eta (t), \label{eq.3}
\end{align} 
where $\eta(t)$ represents white noise with $\langle \eta(t) \rangle =0$, $\langle \eta(t)\eta(t') \rangle =\delta(t-t')$ and $v_0$ represents non-zero constant speed of the particle. Along with the position coordinates $(x,y)$, the particle undergoes a rotational Brownian motion in its orientation $\theta(t)$, controlled by the rotational diffusion constant $D_R$. The "active" nature of the particle is signified by the coupling between position and orientation of the particle. Such models for ABP-s have been used in a broad spectrum of examples including collective motion such as swarming \cite{topaz2004swarming, strefler2008swarming}, flocking \cite{toner2005hydrodynamics}, and clustering \cite{slowman2016jamming, palacci2013living}. Other examples include \textit{E. coli} bacteria tracing the nutrients \cite{viswanathan2011physics}, or sperms going towards the egg \cite{friedrich2008stochastic}. We refer to these excellent reviews for an overview of active transport in complex environments \cite{bechinger2016active, martinez2021active,SR-review}.

\begin{figure}[t]
	\includegraphics[scale = 0.048]{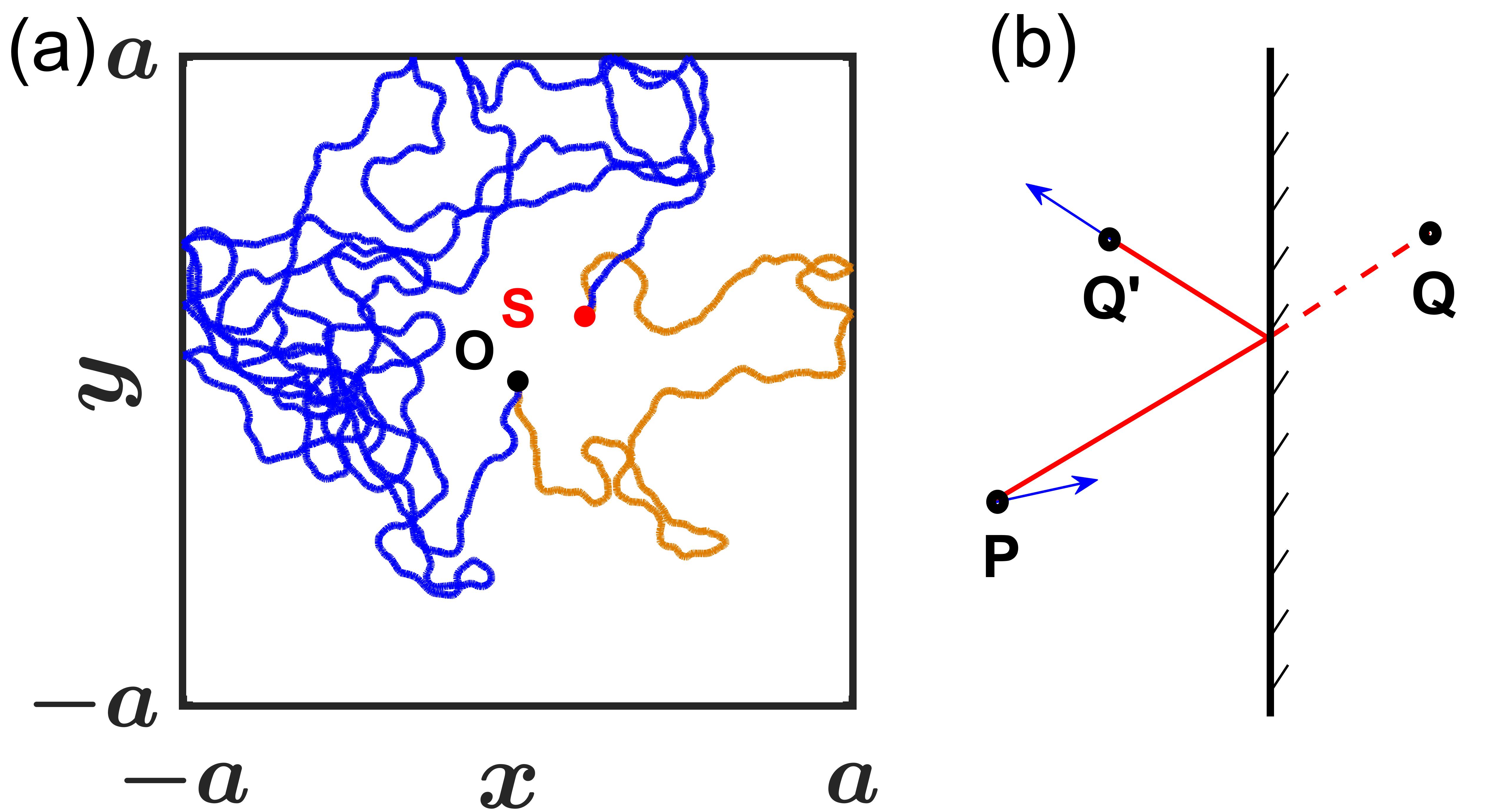}
	\caption{{\bf Typical trajectory of an active Brownian searcher in a confined geometry.} Panel (a): The ABP starts its search process from the source $\rm O$ (black dot) and the process is completed once the target $\rm S$ (red dot) is found.  Here, we portray two such independent paths (indicated by blue and orange curves) traversed by the ABP. Panel (b): When the ABP encounters the boundary on its way, it gets reflected inside the bounded region (for details, see Sec.\ \ref{2-1}). At the time of reflection, the orientation (blue arrow) of the particle changes and gets aligned to the direction of reflection.}
	\label{fig1}
\end{figure}

In this article, our focus is oriented towards the \textit{active search process}. Finding optimal strategy for active particles has been a long standing problem since it is crucial to adopt the right search protocol in complex and crowded environments with scarce resources \cite{benichou2011intermittent,volpe2017topography}. Interestingly, the  topography of the searching process is a determining factor for finding the search time since boundaries, barriers, and obstacles play a crucial role in determining the searcher’s motion \cite{ bechinger2016active,volpe2017topography}. In contrast to a L\'evy walk searcher as considered in \cite{volpe2017topography}, we study motion of an ABP that searches for a target within a homogeneous topography in a confined region in two-dimensional space (see Fig.\ \ref{fig1}(a)).

\subsection{The topography and boundary conditions}
\label{2-1}
We assume that the ABP is confined in a two-dimensional (2D) square region $\mathcal{B} = \{(x,y) \in \mathbb{R}^2 : (x,y) \in [-a,a] \times [-a, a]\}$. The targets ($\rm S$) are assumed to be inside this region and not at the boundaries. Furthermore, the boundaries are considered to be reflective so that there is no loss of probability. In Fig.\ \ref{fig1}(a), we have shown two independent search trials (marked by blue and orange curves) performed by the ABP starting from the source  $\rm O$. The reflecting boundary condition can be understood in the following way (similar to \cite{volpe2014simulation}). Let us imagine that the searcher is at point $\rm P$ at some time instant $t$ inside the boundary and at the next time step $t+\Delta t$, it leaps over the boundary and arrives at a point $\rm Q$ outside the boundary. In such case, we construct a mirror image $\rm Q'$ of $\rm Q$ across that specified boundary so that it is inside the region. Finally, we update the motion of the ABP by assigning its location to $\rm Q'$ at time $t+\Delta t$. In parallel, we also assume that the orientation of the particle has changed along the direction of reflection, which is shown by the blue arrow in Fig.\ \ref{fig1} (b).

\subsection{Target locations} As mentioned earlier, the topography is considered to be a square with diagonal $5\sqrt{2}$.
We furthermore consider one target at a time and measure the search time. The target is located at
different coordinates $\rm A (1,1)$, $\rm B(-2,2)$, $\rm C(-3,-3)$, and $\rm D(4,-4)$ respectively in each such realization. This is to distribute the target over the four quadrants in the 2D plane placed at different distances from the origin. However, due to the homogeneous topography, the search time remains same along the circle centered at the origin and spanned by the target.

\subsection{Initial conditions} In this set-up, we have considered two types of initial conditions. 
Denoting $x_i \equiv x(0),~y_i \equiv y(0)$ and $\theta_i \equiv \theta(0)$ as the initial conditions, we consider the following.
\begin{itemize}
    \item {\bf IC1} (quenched initial condition): We consider fixed position coordinates $x_i=0,y_i=0$, and random initial orientation $\theta_i \in U[0,2\pi]$, where $U[m,n]$ is a uniform random variable between $m$ and $n$.
    \item  {\bf IC2} (annealed initial condition): We consider random position coordinates chosen uniformly from the bounded region $\mathcal{B}$ and the orientation at the initial time is also chosen randomly as before. 
\end{itemize}

\begin{figure*}[hpt]
	\centerline{
		\includegraphics[scale = 0.58]{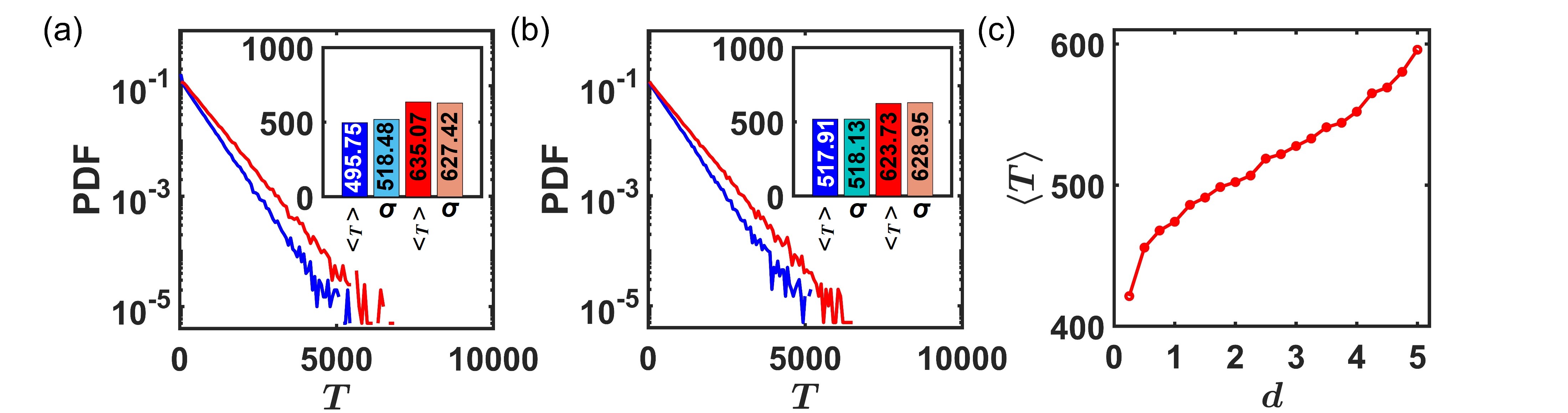}}
	\caption{{\bf First passage time (FPT) densities of the reset free search process and the dependence of MFPT ($\langle T \rangle$) on the distance of the target from the origin.} Panel (a): First passage time densities for the targets $\rm A(1,1)$ (blue), and $\rm D(4,-4)$ (red). Here, we have considered the quenched initial condition where the ABP always starts its search from the origin. The respective mean and fluctuation values are shown in the inset which indicate that $\rm CV>1$ for target $\rm A$, and $\rm CV<1$ for target $\rm D$ (see Table\ \ref{table1} for details). Panel (b): FPT densities for the same targets but for the annealed initial condition when the ABP starts its search from random positions inside the bounded region. In this case $\rm CV$ is always greater than 1 (see inset and Table\ \ref{table1}). Panel (c): $\rm MFPT$ ($\langle T \rangle$) as a function of distance from the origin ($d$) under IC1. $\langle T \rangle$ increases when the distance of the target from origin increases. A circle of radius $d$ centered at the origin is drawn for each of the distances. Ten points on the circle's circumference are chosen which are equally spaced and considered as targets. Then the average of all realizations ($10$ $\times$ $10^6$) is used to calculate $\langle T \rangle$.}

	\label{fig2}
\end{figure*}

\section{\label{sec:level3}First passage time statistics of ABP}
Before studying the effects of resetting on the first passage statistics, it is first instructive to delve deeper into these properties without resetting. Thus, starting from a source, the ABP diffuses inside the bounded region $\mathcal{B}$ until it encounters the target for the first time, thus marking the termination of the search process. The time required for the search completion is random and is often known as the {\it first passage time}. Let, ${\rm S} \equiv(x_{\rm S}, y_{\rm S})$ be the position vector of the target and $X(t) \equiv (x(t),y(t))$ be the position vector of the ABP at time $t$. The target is said to be encountered by the ABP when $||X(t) - \rm S|| < \epsilon$, for some predefined threshold $\epsilon$ ($>0$), which is sufficiently small. We denote this first passage time ($\rm FPT$) by $T$, which is defined as
\begin{equation}
	T = \inf \{t > 0 \; : \; ||X(t) - \rm S|| < \epsilon\}.
	\label{eq.4}
\end{equation}
A wide array of applications in physics, chemistry, biology and other interdisciplinary fields have turned first passage time processes into a long-standing focal point of scientific interest. Fundamental questions include computation of first passage time density, its moments and designing various efficient strategies to make it optimal. We refer to these extensive reviews \cite{benichou2011intermittent,bray2013persistence,redner2001guide} for a detailed overview of the subject.

In what follows we use $f_Z(t)$,  $\langle Z^n \rangle$, and $\sigma(Z)$ to denote, respectively, the probability density function, moment of $n$-th order, and standard deviation of a non-negative random variable $Z$ (e.g., the first passage time $T$ etc.). The corresponding coefficient of variation ($\rm CV$) for the first passage time is defined as
\begin{equation}
	\mbox{CV}  = \dfrac{\sigma(T)}{\langle T \rangle}. 
\end{equation}
Coefficient of variation is a widely used measure of statistical dispersion that infers how some random data points are dispersed around their mean. In other words, it tells us how broad (large fluctuations) or narrow (small fluctuations) a distribution is around their mean.

As our study involves two different initial conditions, we 
investigate these two scenarios for the searching mechanism of the ABP in the following. In both cases, we carry out a comprehensive study for the mean first passage time (MFPT) and fluctuations of FPTs and support the results qualitatively. The results for the FPTs (by performing $10^6$ trials) for two different initial conditions are listed in Table\ \ref{table1}. 

\begin{table}[t]
	\centering
	\caption{Numerically calculated values of mean first passage time ($\langle T \rangle$), standard deviation ($\sigma$), and coefficient of variation ($\rm CV$) for four targets $\rm A$, $\rm B$, $\rm C$, and $\rm D$.  Here we set $D_R=1$, $v_0=1$.}
	\begin{tabular}{|c||p{1.02cm}|p{1.02cm}|p{1.02cm}||p{1.02cm}|p{1.02cm}|p{1.02cm}|}
		\hline Target & \multicolumn{3}{c||} {IC1} & \multicolumn{3}{c|} {IC2} \\
		\cline { 2 - 7 } & $\langle T \rangle$ & $\sigma$ & $\rm CV$ & $\langle T \rangle$ & $\sigma$ & $\rm CV$ \\ \hline 
		$\rm A(1,1)$ & 495.75 & 518.48 & 1.0458 & 517.91 & 518.13 & 1.0004 \\ \hline
		$\rm B(-2,2)$ & 530.79 & 532.82 & 1.0038 & 533.45 & 535.43 & 1.0037\\ \hline
		$\rm C(-3,-3)$ & 568.25 & 566.57 & 0.997 & 561.99 & 566.16 & 1.0074 \\ \hline
		$\rm D(4,-4)$ & 635.07 & 627.42 & 0.9879 & 623.73 & 628.95 & 1.0084 \\
		\hline
	\end{tabular}\label{table1}
\end{table}

\subsection{\label{3:1}Searching with {\bf IC1} (quenched initial condition)} 
\begin{figure}[h]
	\includegraphics[scale = 0.06]{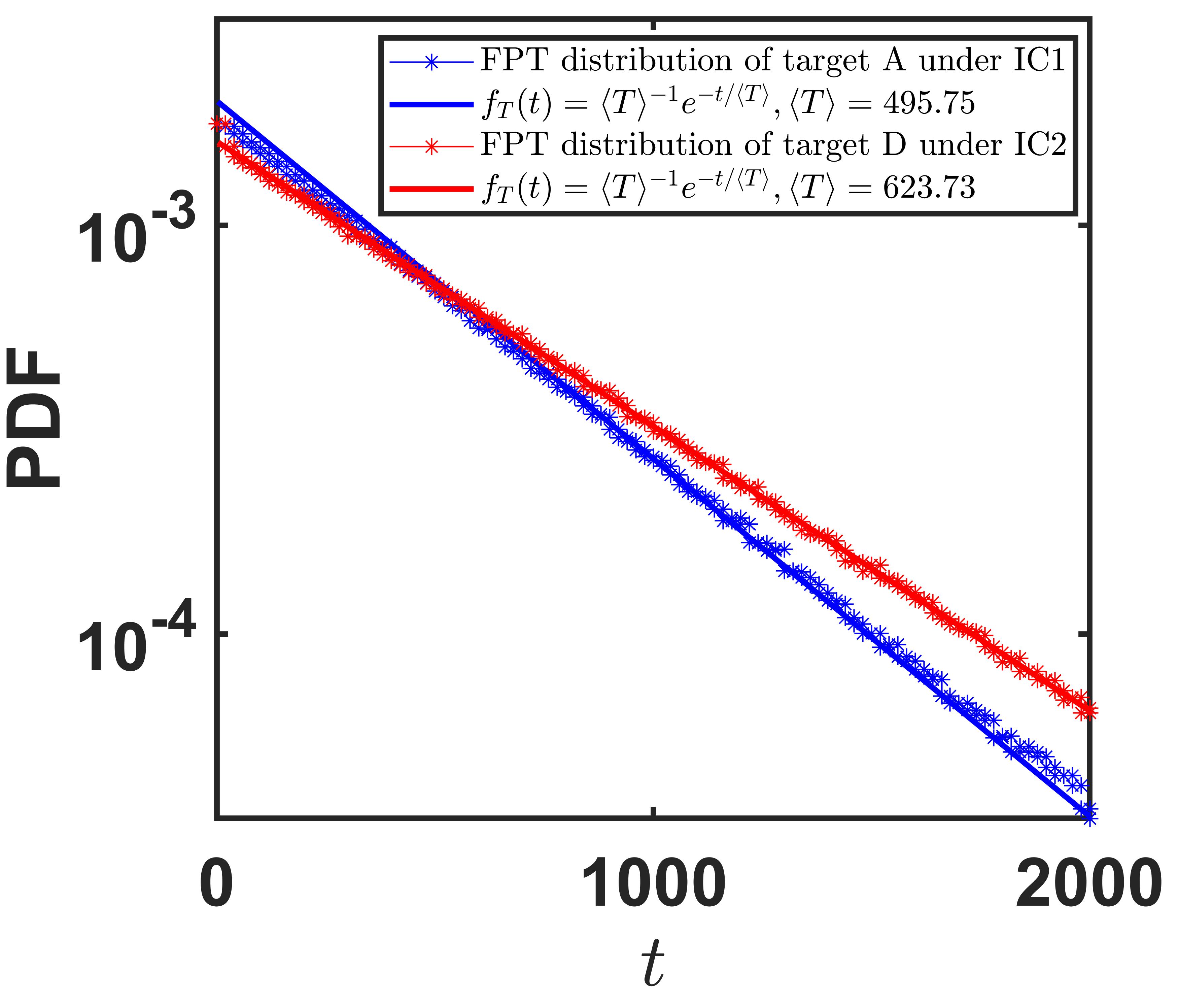}	
	\caption{{\bf Nature of the underlying first passage time densities.} Statistical distributions for the underlying first passage time process for individual target show exponential tails in the asymptotic limit. This observation is found to be robust for both the initial conditions. Blue solid line is the theoretical fit for target $\rm A$ under IC1 where blue stars represent numerical data of first passage times. Similar behavior is observed for target $\rm D$ under IC2 (in red) with a different slope (determined by $\langle T \rangle$).}
	\label{expo}
\end{figure}

We first consider the case when the ABP starts from the quenched initial conditions (IC1) and evolves until it encounters the target. While both $\langle T \rangle$ and $\sigma$ steadily grow as we increase $d$, the distance between the target and the origin, we observe a reduction in the fluctuations $\rm CV$ 
(see Table \ref{table1}). The probability density functions $f_T(t)$ for the first passage time are numerically computed and illustrated for two targets $\rm A$ (blue) and $\rm D$ (red) [Fig.\ \ref{fig2} (a)]. Since the search process is taking place in a finite domain, the first passage time densities, at large time, pertain to exponential tails in contrast to the algebraic tails which usually appear for the search in the unbounded domain \cite{redner2001guide,bray2013persistence}. We refer to Fig.\ \ref{expo} where we have compared the simulation results for the first passage times against the theoretical form $f_T(t)=1/\langle T \rangle e^{-t/\langle T \rangle}$ at large times. Owing to the exponential tails, all the moments of the first passage time remain finite. For brevity, the respective values of $\langle T \rangle$ and  $\sigma$ are provided in the Table. It should be noted that the $CV=\sigma/\langle T \rangle$ may seem exactly unity owing to the exponential distribution, but that will be an incorrect assessment since it will leave out the short and intermediate time contributions which are not exponential.

Clearly, the ABP takes longer time to find the targets that are far away from the source (here $\rm D$) than the ones those are closer to the source ($\rm A$), i.e., $\langle T \rangle_{\rm D}> \langle T \rangle_{\rm A}$ [see Fig.\ \ref{fig2} (c)]. However, the increasing trend of fluctuations are comparatively reduced for a long-range target, and {\rm CV} is dropped below 1 [Table \ref{table1}]. The average of the  $\langle T \rangle$, is plotted against the distance $d$ shown in Fig.\ \ref{fig2} (c). The graphic also shows that when the distance between the target and the fixed source (origin) grows, the ABP takes longer to reach the target.

\subsection{\label{3:2}Searching with {\bf IC2} (annealed initial condition)}
In this subsection, we focus on the motion of the ABP under the annealed random initial conditions. The numerical procedure remains the same and we list the results for $\langle T \rangle$, $\sigma$, and $\rm{CV}$ in Table\ \ref{table1}.
Similar to the previous case, we find that both $\langle T \rangle$ and $\sigma$ increase with the distance. Figure.\ \ref{fig2}(b) showcases the first passage time probability density functions for the targets $\rm A$ (blue) and $\rm D$ (red). The nature of the distribution remains essentially similar to that shown in Fig.\ \ref{fig2}(a). As before the first passage time density for the resetting free process has exponential tails at large times as can be seen from Fig.\ \ref{expo}. 
However, in this case, we find that ${\rm CV}$ is always larger than $1$, i.e., the fluctuations $\sigma$ tend to exceed the mean $\langle T \rangle$.

\begin{figure}[b]
	\includegraphics[scale = 0.035]{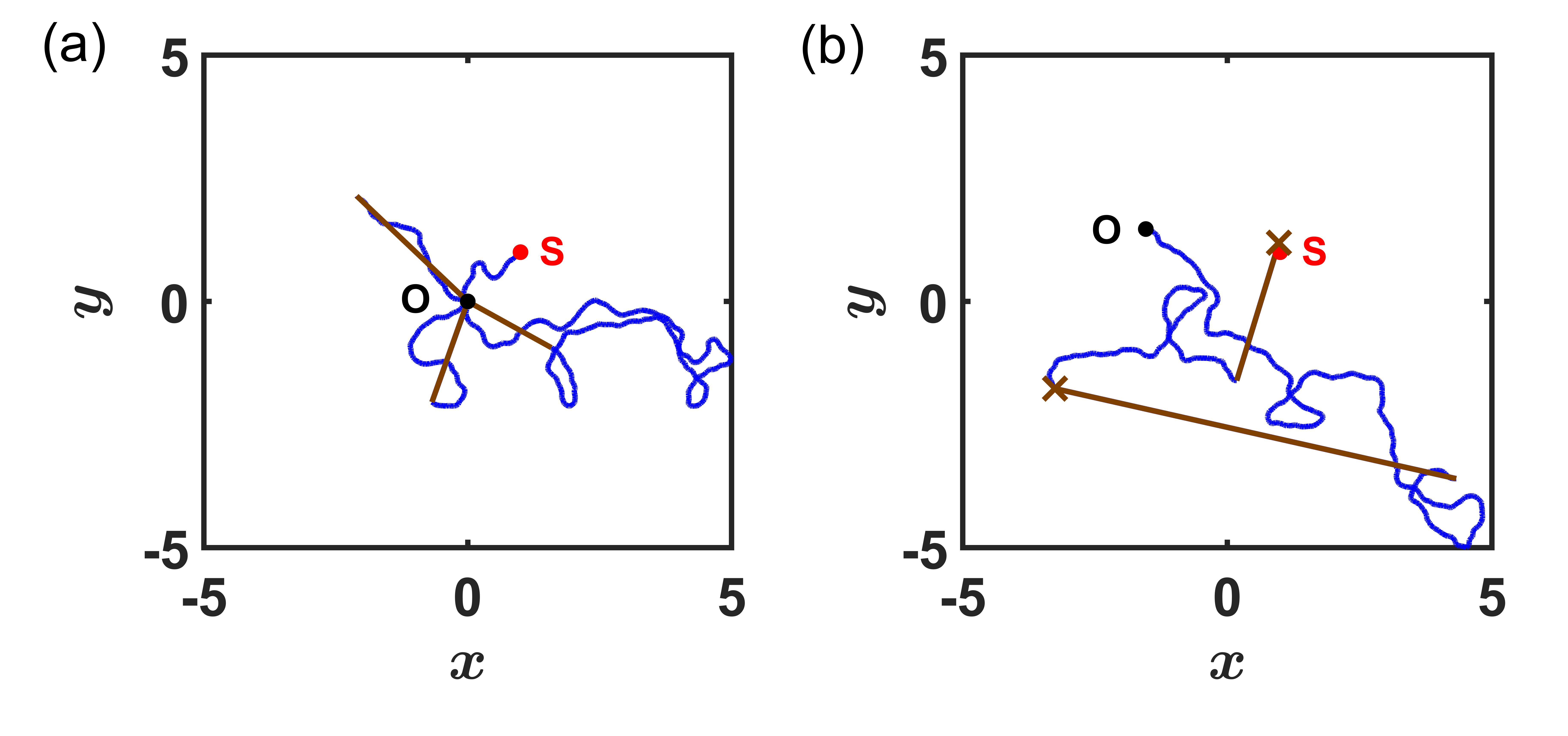}	
	\caption{{\bf Typical trajectory of an ABP subject to stochastic resetting mechanism.} Generically, motion of the ABP is governed by the Eqs.\ \eqref{eq.1}-\eqref{eq.3} until the ABP is stopped and the motion is reset. Between two resetting events, the stochastic motion of the ABP is depicted by blue trajectories and the reset jumps are indicated by brown lines. Panel (a): Resetting protocol I:  After each reset, the particle is reset to the origin $\rm O$. Panel (b): Resetting protocol II: In this protocol, the reset position is completely random and is chosen uniformly from the domain under consideration. In both cases, the target $\rm S$ is placed in the same location. The resetting times for both protocols are drawn from an exponential distribution (Eq.\ \eqref{eq.6}).}
	
	\label{fig3}
\end{figure}
         
\begin{figure*}[hpt]
	\centerline{
		\includegraphics[scale = 0.05]{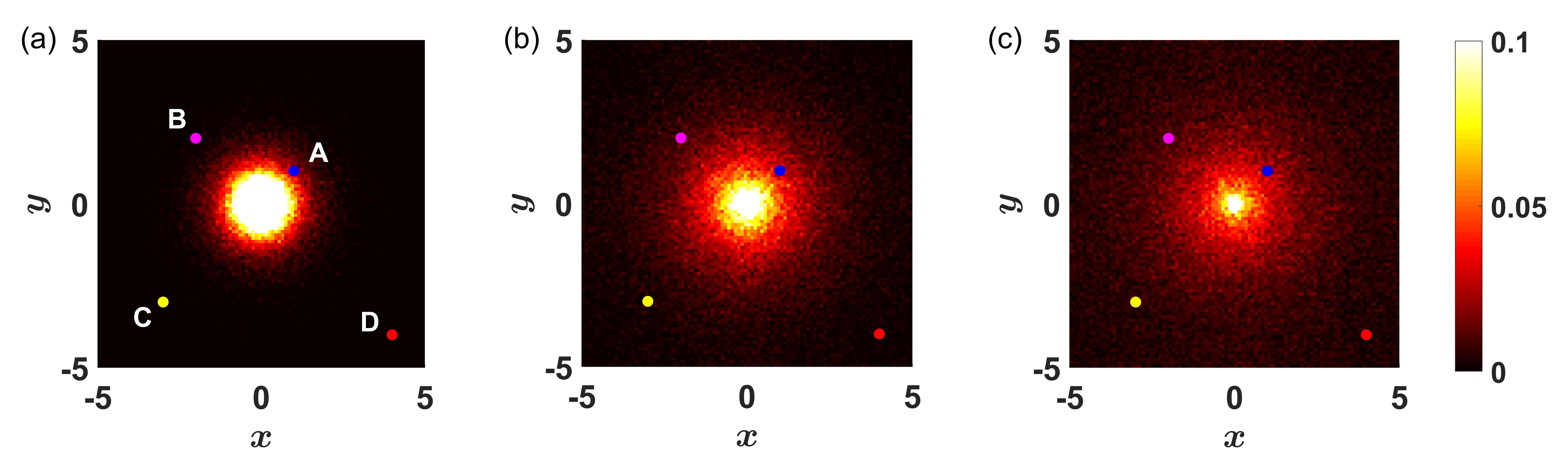}}
	\caption{{\bf Spatial probability distribution of the ABP projected to $(x,y)$ plane under resetting protocol I.} Evidently, the span (excursion) of the phase space trajectory depends on  the mean resetting time $\langle R \rangle$. Here, we have taken three choices for the mean resetting time: (a) $\langle R \rangle = 1.0$, (b) $\langle R \rangle = 5.0$, and (c) $\langle R \rangle = 10.0$. In the colorbar, black represents zero probability density and yellow represents maximum probability. Blue, magenta, yellow, and red dots indicate the location of the targets $\rm A(1,1)$, $\rm B(-2,2)$, $\rm C(-3,-3)$, and $\rm D(4,-4)$, respectively. These targets are juxtaposed externally and do not really hinder the motion of the ABP. Panel (a): For $\langle R \rangle=1.0$, the spatial density is non-zero only within a small radius centered at the origin, within which only A (blue dot) exists. This essentially means that A can be tracked by the ABP  with higher probability while there is almost negligible probability to track the targets C and D.  Panel (b): The intensity of spatial density is more distributed when $\langle R \rangle=5.0$. Since, resetting occurs less frequently, the particle can diffuse more and the spatial distribution is more wider. Naturally, in this case, it is easier to track the targets $\rm A$ (blue dot) and $\rm B$ (magenta dot) who are in more probable zone of the spatial density.  Panel (c): When $\langle R \rangle$ is even further increased to 10.0, the particle can spread even further. The spatial density is maximum at the origin in all the cases as the ABP is reset every time to the origin. However, due to more excursion, all the targets become accessible to the ABP. In the numerical simulation, the probability density is calculated at the time step $10^2$ units. We took $10^5$ realizations of the process (also see Fig. \ref{spread1} for the spread of typical phase space points in one realization) to perform the ensemble averaging.}

	\label{fig4}
\end{figure*}

\section{\label{sec:level4}Resetting mediated First passage time of active Brownian particle}
In this section, we turn our attention to the first passage time statistics of the ABP which is further subjected to stochastic resetting. As mentioned earlier, resetting brings back the ABP to its initial condition after a random time $R$ drawn from a distribution $f_R(t)$. While this distribution could be arbitrary, in here, we assume that the resetting time distribution is exponential such that
\begin{equation}	
	f_R(t) = \langle R \rangle^{-1} e^{- t/\langle R \rangle},
	\label{eq.6}
\end{equation}
where, $\langle R \rangle$ is the mean of this distribution. Simply put, resetting occurs with a rate $\langle R \rangle^{-1}$. Since we do not distinguish between the resetting and initial locations, it allows us to design two different resetting protocols. In what follows, we investigate the effect of resetting on the mean first passage or completion time, denoted by $\langle T_R \rangle$. For brevity, the resetting mechanisms are illustrated schematically in Fig.\ \ref{fig3} and discussed further in the following subsections. While passing, we also note that there have been some recent works on active systems such as ABP (like here), and RTP under stochastic resetting \cite{evans2018run,kumar2020active,santra2020run,abdoli2021stochastic,abdoli2021stochastic}; 
however these works deal with relatively simple geometry \& target configurations, fixed resetting conditions and external fields which in turn renders many exact results. In stark contrast, our work is motivated by realistic topography \cite{volpe2014simulation,volpe2017topography} and an arbitrary choice of targets -- although it limits the tractability of exact results, it showcases a realistic scenario and furthermore, the effect of environmental complexities and initial conditions is explored in details.

\subsection{Designing resetting protocol I -- quenched initial conditions}
\label{Resetting Protocol 1}

The first resetting protocol is associated with {\bf IC1}, i.e., the quenched or fixed initial conditions. Here resetting is performed on the particle in a way which brings it back to the same configuration (IC1) instantaneously. Fig.\ \ref{fig3} (a) depicts this scenario where the brown lines indicate the occurrence of resetting throughout the journey. The time interval between two successive resetting follows Eq.\ \eqref{eq.6}. The FPTs of the specified target(s) depend on the rate at which resetting occurs.

Before getting into the first passage properties, we first investigate the spatial properties for
the active Brownian motion under resetting (i.e., by scanning the phase space coordinates of the ABP up to a given time -- see Fig. \ref{spread1} and then performing an ensemble average) in the absence of target(s). To be precise, we observe the spatial distribution of the ABP under resetting in the $(x,y)$ plane ( Fig.\ \ref{fig4}). We plot the spatial distributions for three alternative values of $\langle R \rangle$, as illustrated in Fig.\ \ref{fig4}. When $\langle R \rangle=1.0$, the ABP is frequently reset to the origin  and the particle's trajectory is constrained to a tiny circular region (yellow area) around the origin for the majority of the time, as seen in Fig.\ \ref{fig4} (a). As we increase $\langle R \rangle$, resetting occurs less frequently which allows the ABP to cover a larger area from the source (which is the origin in this case). Consequently, the accessible region is broader than before which is clear from Fig.\ \ref{fig4} (b)-(c) for $\langle R \rangle=5.0$ and $\langle R \rangle=10.0$, respectively.

  \begin{figure}[hpt]
 	\centerline{
 		\includegraphics[scale = 0.055]{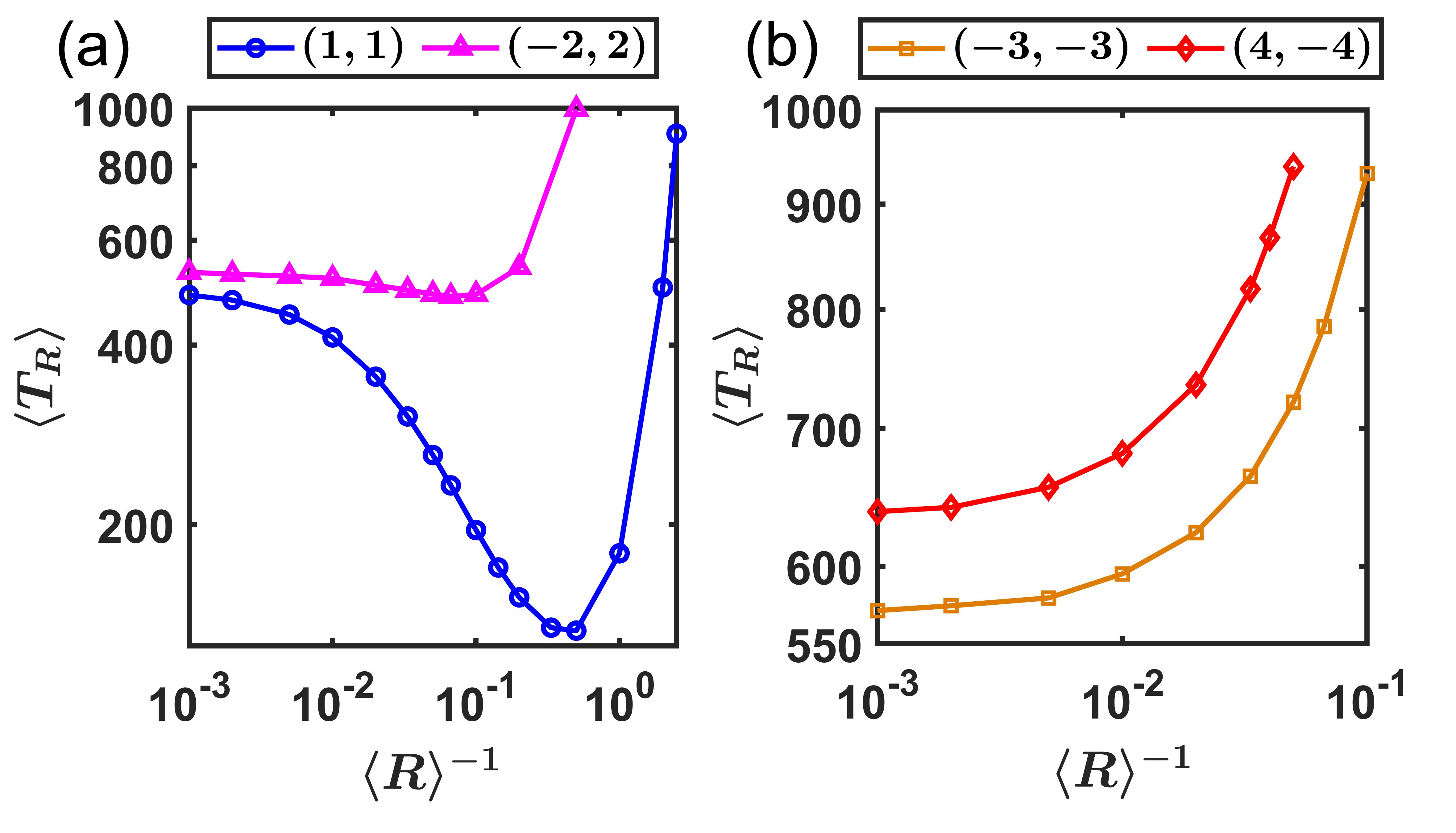}}
 	\caption{{\bf Resetting protocol I (when the ABP is always reset to the origin).} Plot of $\langle T_R \rangle$ as a function of the resetting rate $\langle R \rangle ^{-1}$. Small $\langle R \rangle^{-1}$ corresponds to the almost no-resetting case (i.e., the limit  $\langle T_R \rangle \approx \langle T \rangle$) while large $\langle R \rangle^{-1}$ implies frequent resetting. Panel (a): optimality in $\langle T_R \rangle$ is observed for the targets $\rm A (1,1)$ and $\rm B (-2,2)$. For target $\rm A$, the optimal value 
  of $\langle T_R \rangle$ is	$192.8$ ($< 495.75$) which is achieved 
 	for $\langle R \rangle^{-1} \approx 0.5$. For target $\rm B$, the optimality is achieved at $\langle R \rangle^{-1}  \approx 0.1$ and  $\langle T_R \rangle$ is $483.5$ ($< 530.79$). Panel (b): Resetting is not beneficial for the other two targets $\rm C$ and $\rm D$ as $\langle T_R \rangle$ is larger than $\langle T \rangle$ for all values of $\langle R \rangle^{-1} $. We perform numerical simulations by taking $10^6$ realizations for each value of $\langle R \rangle^{-1}$.}
 	\label{fig5}
 \end{figure}

We now turn our attention to the mean search time in the presence of resetting. We summarize our results in Figure  \ref{fig5} where we demonstrate the variation of $\langle T_R \rangle$ as a function of the resetting rate $\langle R \rangle^{-1}$. For relatively small values of $\langle R \rangle^{-1}$, resetting is quite infrequent and thus $\langle T_R \rangle \approx \langle T \rangle$ (i.e., the mean first passage time of the free ABP). For target $\rm A$ (blue), $\langle T_R \rangle$ starts decreasing with the increment of $\langle R \rangle^{-1}$ until it reaches a minimum at $\langle R \rangle^{-1} \approx 0.5$ (i.e., $\langle R \rangle  \approx 2.0$) (see Fig.\ \ref{fig5}(a)). This can also be perceived from the spatial distribution in Fig.\ \ref{fig4} (a) where the target $\rm A$ (blue dot) lies very close to the high spatial density zone for $\langle R \rangle = 1.0$ and thus it can be accessed easily by the ABP. From this optimal, $\langle T_R \rangle$ increases with $\langle R \rangle^{-1}$. For the target $\rm  B$ (magenta), the optimality occurs at $\langle R \rangle^{-1}  \approx 0.1$. However, optimal $\langle T_R \rangle$  for the target $\rm B$ is higher than that of $\rm A$, as the spatial intensity slowly decreases throughout the space if $\langle R \rangle$ is  gradually  increased (i.e., $\langle R \rangle^{-1}$ is decreased). A systematic reduction of $\langle T_R \rangle$ for targets which are far away from the origin is not possible (see yellow and red curves in Fig.\ \ref{fig5}(b)). So, it appears that resetting is advantageous for reduction of the mean first passage time if the targets are located at a specific region of $\mathcal {B}$ (particularly, close to the origin in this case). In what follows, we will demonstrate how one can improvise on this strategy to get hold of the targets that are far away from the origin as well making the search process more efficient in general.

\subsection{Designing resetting protocol II -- annealed initial conditions}
\label{Resetting Protocol 2}
One way of overcoming the limitation of protocol I might be to design a resetting mechanism which would treat all the points inside $\mathcal{B}$ in the same footing, rather than being biased to any particular region. This indication leads us to work with annealed initial conditions (\textbf{IC2}) where the initial position and orientation of the ABP are chosen at random. Now, a new resetting protocol can be implemented which resets the ABP's location to any randomly chosen point inside $\mathcal{B}$. In this way, the statistical identical behavior of the initial and resetting locations is maintained. In Fig.\ \ref{fig3} (b), this resetting mechanism is portrayed where the brown lines represent reset paths and the brown crosses indicate the points where the particle is reset. In contrast to the resetting protocol I, here resetting may send the particle very close to the target, independent of the target's location. This causes a significant reduction to the first passage times ($T_R$) and eventually to their mean ($\langle T_R \rangle$).

\begin{figure}[hpt]
	\centerline{
		\includegraphics[scale = 0.072]{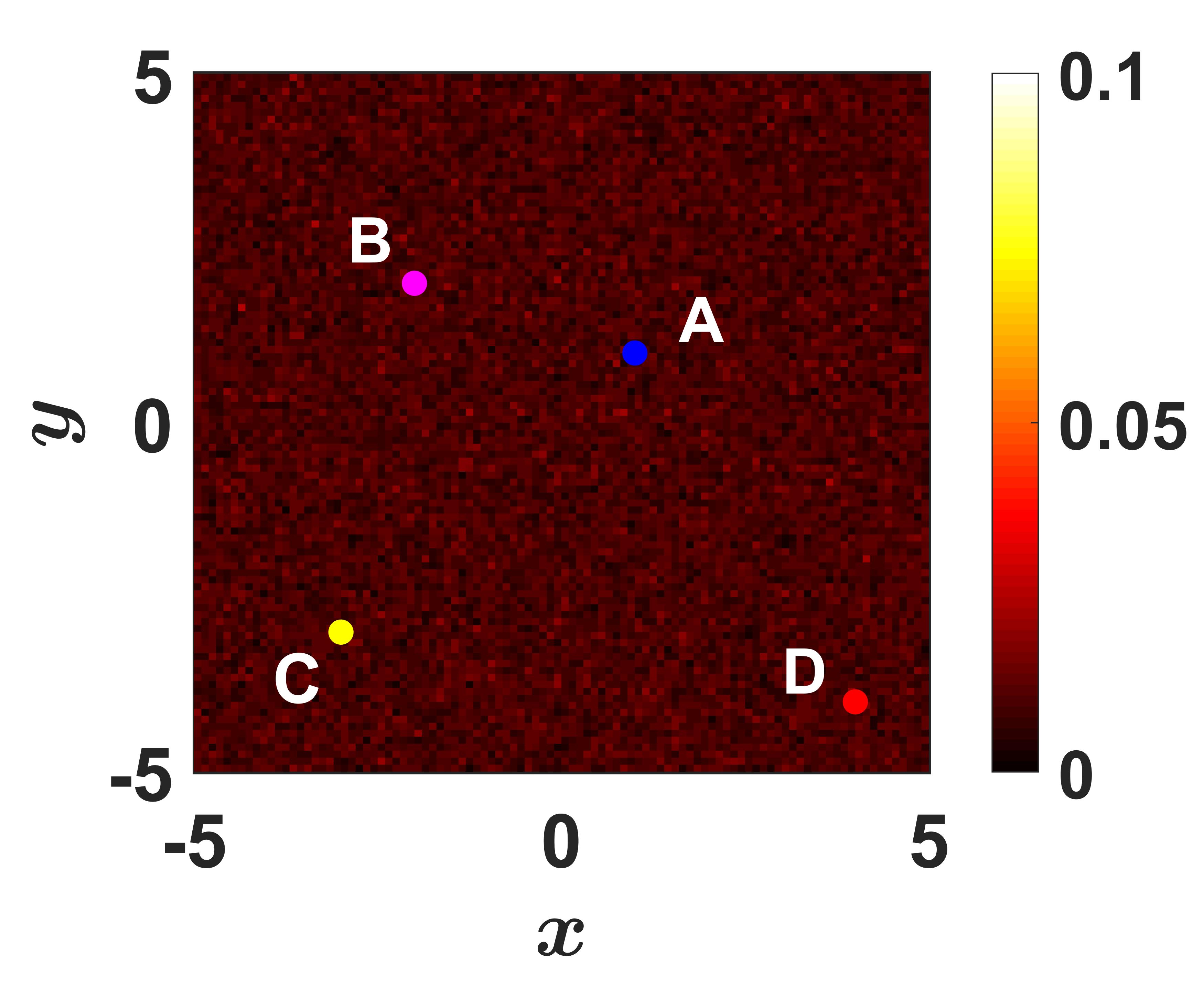}}
	\caption{{\bf Spatial probability distribution of the ABP projected to the phase space under resetting protocol II.} Here the particle is reset to random positions (uniformly) inside the region. We have chosen $\langle R \rangle=5.0$ for this spatial probability density plot. However, the density plot shows same qualitative behavior for any other values of $\langle R \rangle$. Since the particle is reset to random positions, it naturally spans the whole region and the spatial density is almost uniform as evident from the figure. Hence, the targets $\rm A$, $\rm B$, $\rm C$, and $\rm D$ are almost equally accessible by the ABP. Note that, as before, the targets are superimposed externally (in other words, the simulation is done for the target free process) and they do not really affect the motion. In the numerical simulation, the probability density is calculated at the time step $10^2$ units. We took $10^5$ realizations of the process (also see Fig. \ref{spread2} for the spread of typical phase space points in one realization) to perform the ensemble averaging.} 
	\label{fig10}
\end{figure}

\begin{figure}[hpt]
	\centerline{
		\includegraphics[scale = 0.07]{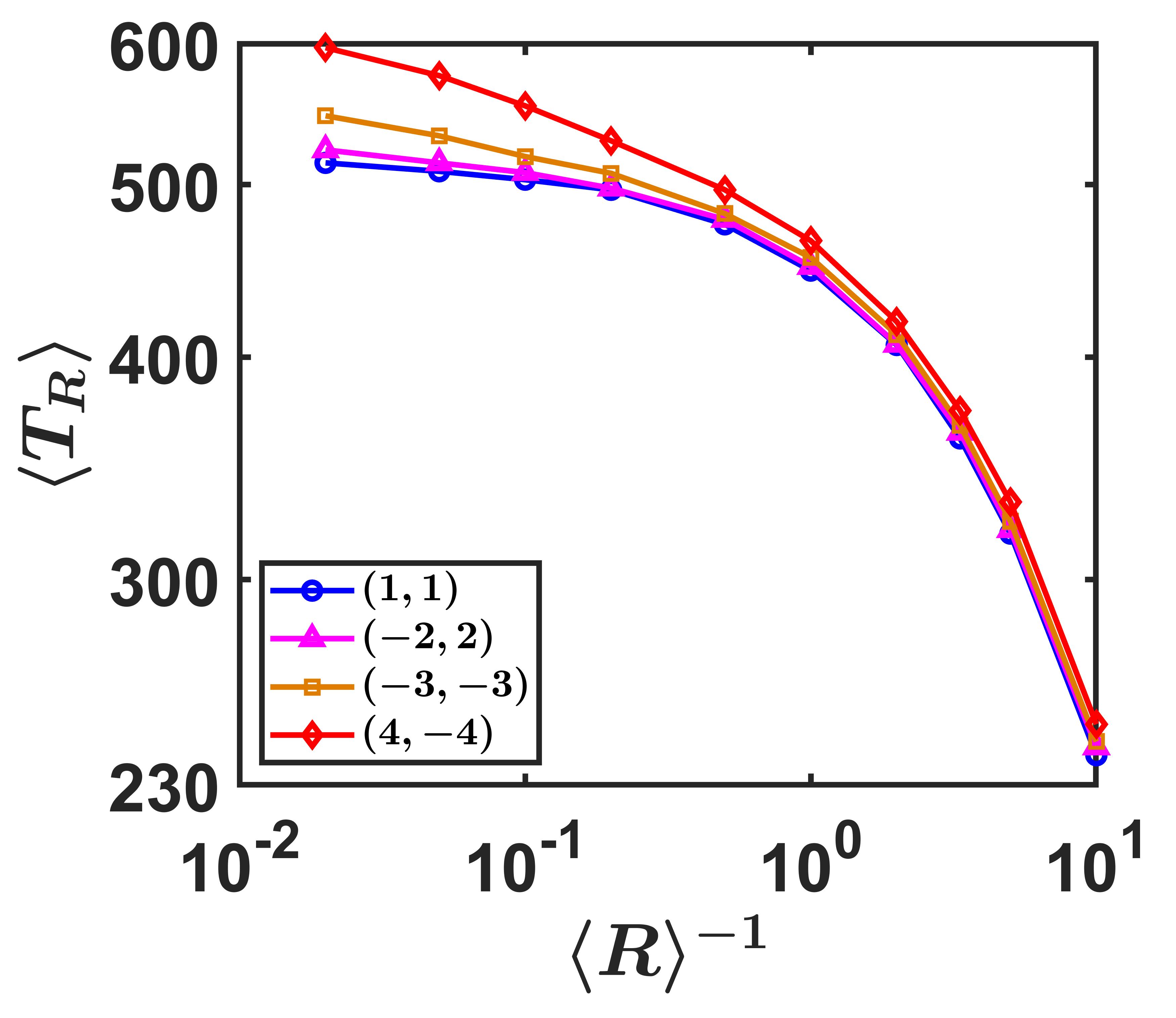}}
	\caption{{\bf Resetting protocol II (when the ABP is reset to random positions).} Plot of $\langle T_R \rangle$ as a function of $\langle R \rangle^{-1}$ when the particle is reset to random positions under resetting protocol II. When $\langle R \rangle^{-1}$ is large, the particle is reset frequently to random positions and it soon spans the whole space in relatively small time. As a result, $\langle T_R \rangle$ tends to decrease as we increase  $\langle R \rangle^{-1}$. The mean first passage time plots for all the targets show the same qualitative behavior when $\langle R \rangle^{-1} \gtrapprox 10^{0}$ (frequent resetting).}
	\label{fig6}
\end{figure}

As before, to gain a better insight, we first look into the spatial distribution of the ABP (see Fig.\ \ref{fig10}) in the above-mentioned configuration but excluding the targets. It is easy to see that the ABP covers the whole region $\mathcal{B}$ uniformly for $\langle R \rangle = 5.0$ (also see Fig. \ref{spread2} where we have shown 
all the phase space points up to a given time in one realization). The nature of this distribution remains invariant under the change of $\langle R \rangle$, i.e., for any values of $\langle R \rangle$, the entire territory is almost uniformly accessed by the ABP. Since the target(s) are any point(s) within this region, the search time under this condition also remains invariant. This emerges as a notable feature for this strategy.

Keeping this wisdom in mind, we now proceed to study the variation of $\langle T_R \rangle$ as a function of the resetting rate $\langle R \rangle^{-1}$. At large $\langle R \rangle^{-1}$, resetting occurs frequently which allows the ABP to travel from one random location to another very rapidly. This increases the probability of finding the target as well as reduces the time for doing so. Consequently, the value of $\langle T_R \rangle$ is remarkably smaller for large $\langle R \rangle^{-1}$ compared to resetting protocol I. This is demonstrated in the $\langle T_R \rangle$ versus $\langle R \rangle^{-1}$ plots in Fig.\ \ref{fig6}. As predicted before, the mean search time for all the targets more or less have the same qualitative behavior and for large $\langle R \rangle^{-1}$, the $\langle T_R \rangle$ values are almost identical. In particular, we note that $\langle T_R \rangle$ (for all the targets) is reduced by almost three fold in comparison to the resetting free process (see the Table \ref{table1} for comparison). Interestingly, implementation of this protocol reveals that there is an optimal rate at $\langle R \rangle^{-1} \to \infty$ for which the search time is the \textit{lowest} for all the targets. This is in accordance with our general prediction made in the previous paragraph. Such instances can also be found in \cite{besga2020optimal}.

\section{\label{sec:level5}Theoretical developments}
So far, most of our analysis have been computational in nature owing to the intractability of active Brownian particles in a confined topography in the presence of targets. In this section, we develop an invasive method borrowing ideas from the general framework of first passage under restart. Although exact expressions can not be obtained for such model systems, this semi-analytic approach turns out to be a useful method for computing the first passage time statistics that is further corroborated with direct numerical simulations \textit{\`a la} Langevin. In what follows, we first explain the effects of resetting and why indeed resetting expedites the active search process. We then provide a semi-analytical estimation of the mean first search time under resetting and compare with numerical simulations.

\subsection{Large search time fluctuations of an ABP and effect of resetting}
What determines the average search time of an ABP in such scenario? The answer to this question clearly depends on the diffusivity, initial configuration, overall topography, and coordinates of the target location. Naturally, these parameters can affect the search process causing in large stochastic fluctuations in the search time making it an inefficient process. It is thus crucial to design strategies that could mitigate fluctuations induced search delays. The theory of first passage time under restart teaches us that a simple resetting mechanism can reverse the deleterious effects of large fluctuations in arbitrary search processes \cite{pal2017first,pal2022inspection}, thus making them more efficient. This property has since been demonstrated in numerous simple diffusive and non-diffusive processes \cite{evans2020stochastic}. In here, we investigate this property for the system of an ABP in the aforementioned geometry.

\begin{figure}[hpt]
		\centerline{
			\includegraphics[scale = 0.37]{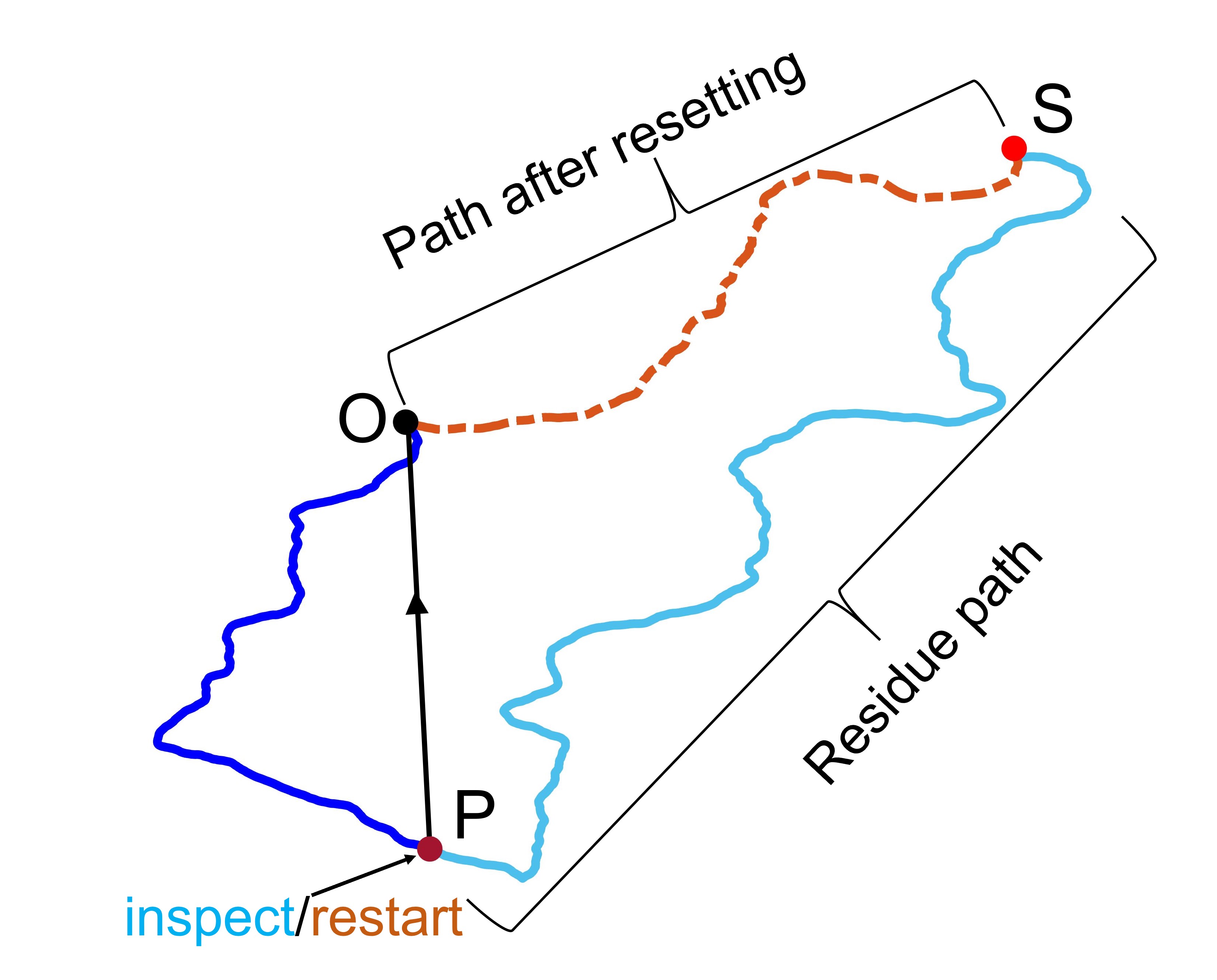}}
		\caption{{\bf A trajectory based analysis for the inspection and restart mechanism.} The ABP begins its search from $\rm O$ (black dot) for the target $\rm S$ (red dot) along the blue-colored trajectory. At a random point in space, say $\rm P$ (maroon dot), motion of the ABP is either inspected or restarted. In the former case, the motion of the ABP remains uninterrupted and we just inspect the remaining/residual path (cyan trajectory) and its corresponding traverse time ($T_{\rm residual}$) till it reaches the target $\rm S$. In the second case, the ABP is reset to the source point $\rm O$ from the point $\rm P$ from which it restarts the search process (shown by the red dashed trajectory). In the case of almost no-resetting afterwards, this time  should be $T $ (see the main text). Resetting will expedite the search process iff, on average, $\langle T \rangle <  \langle T_{\rm residual} \rangle$.}

		\label{fig9}
	\end{figure}

To simplify the problem, let us assume that the ABP starts from a source (say, $\rm O$) and searches for a single target (say, $\rm S$). In the schematic Fig.\ \ref{fig9}, the source ($\rm O$) and the target ($\rm S$) are represented by black and red dots, respectively. To proceed further, we will first analyze the search process without resetting mechanism and then compare with the resetting induced process. This should be done from the reference frame of an external observer who randomly arrives at some point along the trajectory and either inspects (which is the case in the absence of resetting) or resets (naturally in the presence of resetting) the ongoing process. For the former case, recall that the time taken by the ABP to find the target from the source is denoted by $T$ -- the mean being $\langle T \rangle$. However, the external observer may arrive at a point (say maroon dot point $\rm P$ in Fig.\ \ref{fig9} which is different than the source point $\rm O$) while the process has already started. Although the observer will not interrupt the process, the time remaining for the process completion would be a different random variable (denoted by $T_{\rm residual}$) pertaining to a different distribution. The trajectory colored in cyan in Fig.\ \ref{fig9} corresponds to this path, called as the residue path and the waiting time for this observer is often known as the residual time, with the following form for its mean \onlinecite{pal2022inspection}
	\[\langle T_{\rm residual} \rangle = \dfrac{\langle T^2 \rangle}{2\langle T \rangle}.\]

	\par In the second scenario, the observer arrives at this arbitrary point $\rm P$ and resets the particle to the source $\rm O$ from the point $\rm P$ (indicated by black line)
and enforces to restart the search. The path traversed by the ABP after being reset is represented by a red dashed trajectory in Fig.\ \ref{fig9}. Following restart, the average time taken by the ABP to reach the target from the source again would be simply
$\langle T \rangle$ 
since it can be considered as a new search trial. Furthermore, this assumes the limit to an infinitesimal resetting rate so that hardly one resetting event can be experienced. Comparing the two scenarios explained above, one can assert that resetting expedites the search time if the average completion time following restart is less than the mean residual time, i.e.,
\begin{align}
	\langle T \rangle < \langle T_{\rm residual} \rangle &= \dfrac{\langle T^2 \rangle}{2\langle T \rangle},
\end{align}
which, after simplification, yields
\begin{equation}
	{\rm CV} > 1,
	\label{cv}
\end{equation}
where recall that $\rm  CV=\frac{\sigma}{\langle T \rangle}$ is the coefficient of variation for the search time in the absence of resetting. $\rm  CV$ is a measure of statistical dispersion that quantifies how broad a distribution is. Thus the above result signifies that resetting is helpful in the reduction of the mean search time only when the uninterrupted search process has large fluctuations (namely  $\rm CV >1$) \cite{pal2017first,pal2022inspection}.

The set-up of the ABP can be used as a testing ground for this criterion. For the quenched initial condition IC1 when the ABP always starts its search from a fixed location (Sec.\ \ref{3:1}), we find that the criterion is satisfied for the targets A and B while it does not hold for the targets C and D. For given $D_R=1$, Table\ \ref{table1} summarizes the values of $\rm CV$ collected numerically from the data set of first passage time for individual targets $\rm A$, $\rm B$, $\rm C$, and $\rm D$. This simply states that resetting is useful for the targets that are close to the initial locations where the underlying search times have $\rm CV>1$ (see Table\ \ref{table1} and Fig.  \ref{fig5}(a)). The diffusion mediated search gradually becomes more inefficient as the ABP looks for targets that are far away. This results in larger mean and fluctuations but somewhat bounding the $\rm CV$ below 1 (see Table\ \ref{table1}). Fig.  \ref{fig5}( b) corroborates with this observation as resetting hinders (disobeying Eq.\ \eqref{cv}) the completion in such cases.

This behavior can also be understood qualitatively from the spatial probability density plots in Fig.\ \ref{fig4}. The plot essentially shows the accessibility of the targets for different values of resetting rate. It is clear that the targets $\rm A$ and $\rm B$ always belong to high probability region, and thus can be detected easily by the ABP. On the other hand, the detection probability for the targets $\rm C$ and $\rm D$ remains quite low for a finite range of $\langle R \rangle$. This is rather evident since resetting occurs persistently to the origin and thus the local region around the origin becomes more accessible by the ABP. This reasoning confers with the numerical results as we have discussed above. Note, however, that for random resetting locations, this argument does not hold as we will show below.

In the case of annealed initial condition, the ABP starts its search from a random point inside the region $\mathcal{B}$. This point is statistically drawn uniformly within the range of the area. Following reset, another random point is chosen uniformly and the particle is teleported back there from where it restarts its search. In this case, we find that the criterion $\rm CV>1$ is quite robust for all the targets. Table\ \ref{table1} summarizes the values of $\rm CV$ of the underlying first passage time for the targets $\rm A$, $\rm B$, $\rm C$, and $\rm D$ which clearly shows that the randomness in the initial condition makes the search more costly resulting in large fluctuations, i.e., $\rm CV>1$. Naturally, these conditions are ideal for resetting to work and we find an overall reduction in completion time (see Fig.\ \ref{fig6}).

As before, this behavior can again be underpinned by taking the route to the heatmaps for the spatial probability density as shown in Fig. \ref{fig10}. Since the initial as well resetting locations are fully randomized over the entire geometry, there is no as such priority for the target locations. In other words, all the targets are equally accessible to the ABP as the spatial probability is spanned uniformly. Thus, resetting works uniformly irrespective of the target locations.

The discussion so far has been focused on a fixed rotational diffusion constant ($D_R$). However, by suitable modulation of $D_R$, fluctuations can be even enhanced further thus resulting in larger coefficient of variation ($\rm CV$). We discuss these aspects in details in Appendix\ \ref{appendixa}.

 \begin{figure*}[hpt]
	\centerline{
		\includegraphics[scale = 0.045]{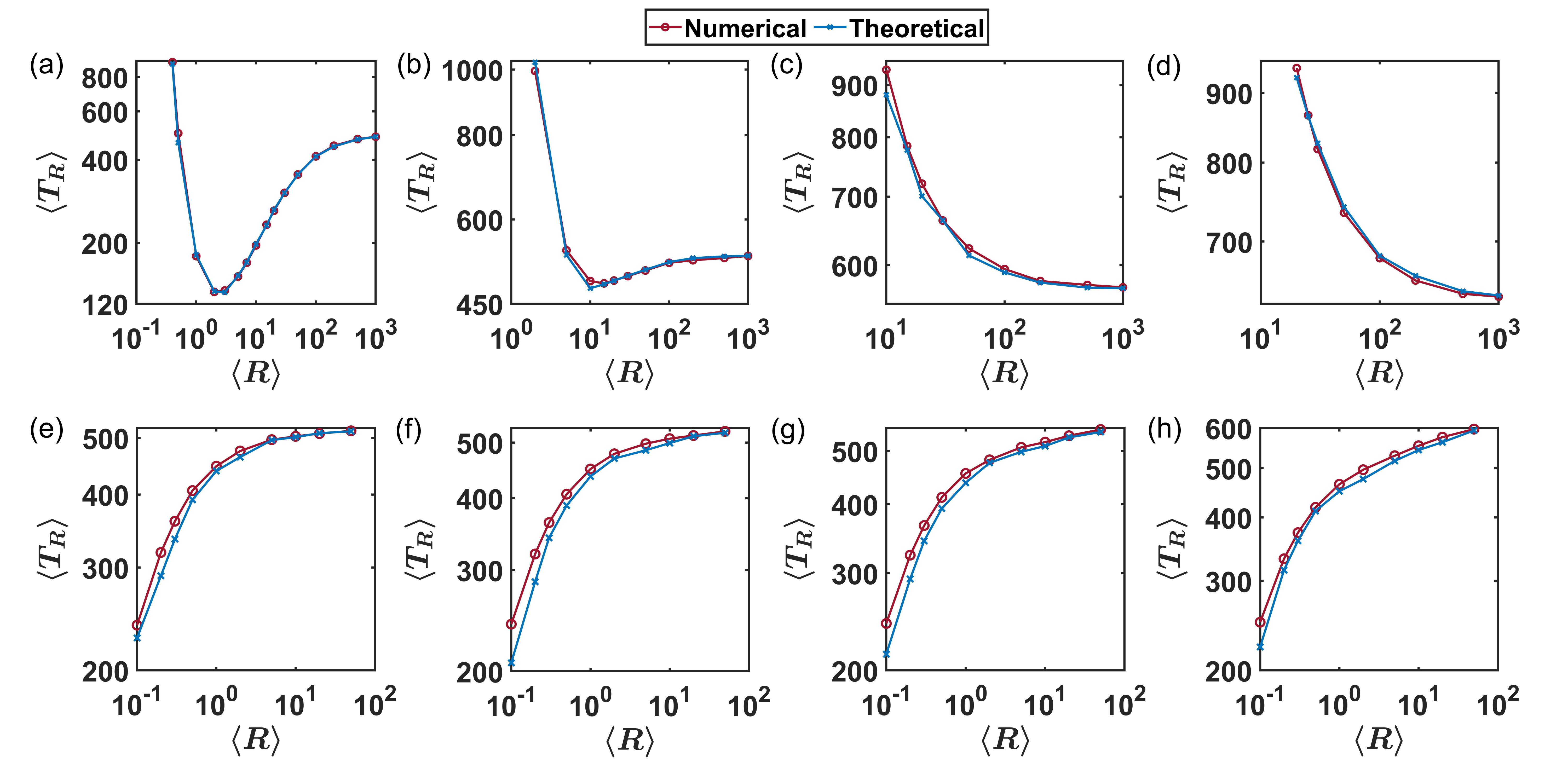}}
	\caption{{\bf Comparison between the semi-analytical and numerical simulations of $\langle T_R \rangle$ for resetting protocol I and II.} In all the cases, red markers indicate numerical values of $\langle T_R \rangle$ calculated from stochastic simulations. On the other hand, blue crosses mark theoretical values of $\langle T_R \rangle$ obtained from using Eq.\ \eqref{eq.9} and the method mentioned in the main text. Numerically each point is a mean of $10^6$ realizations. Panels (a)-(d): $\langle T_R \rangle$ values with varying $\langle R \rangle$ under resetting protocol I. Panels (e)-(h): the same under resetting protocol II. Panels (a) and (e) are for the target $\rm A$, (b) and (f) are for target $\rm B$, (c) and (g) are for target $\rm C$, (d) and (h) are for target $\rm D$. We observe an excellent agreement between the theory and numerical simulations.}
	\label{fig8}
\end{figure*}

\subsection{Mean search time of ABP under resetting} We now turn our attention to the estimation of the mean search time in the presence of resetting. Recently, a formalism namely ``First passage under restart'' was proposed to compute the mean search time of an arbitrary first passage process which is further subjected to generic restarts. Power of this approach should be appreciated since it does not depend on the specific nature of the underlying dynamics. For brevity, we will briefly sketch out this method fitted appropriately to our context.

We begin by reminding ourselves that the random search time in the absence of restart is denoted by $T$. Thus, if the ABP starts from some initial conditions, and finds the target in a random time $T$ without any interruptions, the process ends. Suppose now that the motion of ABP is stopped and compelled to restart from the initial configuration after some random time $R$. In such case, the ABP renews its search. The process is repeated until the ABP finds the target. Letting $T_R$ 
be the random completion time of this compounded search process (and accounting for the above-mentioned possibilities), one can write

\begin{equation}
	T_R = \begin{cases}
		T, & \text{if $T$ < $R$}\\
		R + T_{R}^\prime, & \text{if $R \le T$}
	\end{cases}
\label{eq.8}
\end{equation}
where $T_R^\prime$ 
is a random variable which is independent and identical to
 $T_R$. Using Eq.\ \eqref{eq.8} and taking expectations we can find \cite{pal2017first}

\begin{equation}
	\langle T_R \rangle = \frac{\langle \min(T, R) \rangle}{{\rm Pr} (T<R)},
	\label{eq.9}
\end{equation}
where $\min(T, R)$ is a new random variable and ${\rm Pr} (T<R)$ is the probability that $T<R$. Notably, both these quantities can be computed from the distributions of 
$T$ and $R$. Although there exists formal expressions for these quantities, exact results can be obtained only for simplified cases. The model for ABP does not provide analytically tractable close expressions for first passage time distribution $f_T(t)$, thus we will evaluate Eq. (\eqref{eq.9}) numerically from the data series obtained for $T$ (already in disposal from Sec.\ \ref{sec:level3}) and $R$ (externally employed) respectively. If the data series contains $X$ number of inputs for $T$ and $R$, one can then compute $\min(T, R)$ and mark yes/no for $T<R$ (/otherwise) for each entry $i=1,2,\cdots X$. Finally, the ensemble average is done over these entries (see Github link \cite{sar2022} for more details).

Following the semi-analytical procedure stated above, we calculate the mean search time of the ABP under resetting using Eq.\ \eqref{eq.9}. First passage times for the target are collected by taking $10^6$ trials for both kinds of sources. The mean first search times are plotted in Fig.\ \ref{fig8} for different values of $\langle R \rangle$. Recall that these quantities were also calculated using direct Langevin simulations in Sec\ \ref{sec:level4}. Figure\ \ref{fig8} demonstrates a comparison between the results obtained from pure numerical simulations and that obtained via the semi-analytical approach. In the upper panel of Fig.\ \ref{fig8}, we present the results for mean search time under resetting protocol I. While the red circles represent search times that are the numerically obtained, the blue markers (crosses) represent the same obtained semi-analytically. Similarly, the lower panel of Fig.\ \ref{fig8} showcases a comparison between the numerical and semi-analytical results for search times under resetting protocol II. In both cases, we find the results to be in excellent agreement with each other. This study confers the power of this invasive approach where one can make progress with a hand-to-hand intuitive numerical analysis along with the theory.

\section{Discussion and future perspectives}
Study of first passage processes has always been a fascinating research topic in many interdisciplinary fields. Any dynamical process that starts from a well defined initial configuration and looks for designated or sparse search locations can broadly be classified as FPT process. In statistical physics, a myriad of basic questions and properties have been investigated for passive searchers such as simple diffusing particle. However, searchers in real life are ``active'' driven in non-equilibrium conditions. These searchers usually transport cargos or other important constituents from one place to other. Naturally, active transport is a ubiquitous phenomena that depends non-trivially on the environmental complexities such as geometric constraints, mechanical cues, chemical gradients, and hydrodynamic fluid flow \cite{bechinger2016active,martinez2021active}. Designing various transport and search strategies is an important challenge for the events to take place in a timely manner. Our work proposes one such strategy for active search. Here, we design an intermittent resetting based search strategy for an active system which consists of a single active Brownian particle in a confinement in the presence of targets that require detection for the process completion.

Our work shows that a suitable implementation of resetting can result in a significant reduction of the mean search time. This observation has been found quite robust under various resetting protocols: namely quenched (fixed) condition when the ABP is always returned back to the same place, and annealed (fluctuating/random) condition when the ABP is reset back to random positions. Our study unravels that the \textit{randomized resetting protocols} are more efficient than the quenched resetting protocol in expediting generic search processes. Using a semi-analytical approach, we have checked our numerical results with the analytical results which show an excellent agreement. In particular, we examine a criterion that puts a bound on the relative fluctuations of the first passage time for the resetting free process. The criterion asserts that resetting works in favour if the underlying search process has significant large fluctuations around the average first passage time. Thus, resetting based search offers a generic advantage when the 
underlying active search process faces uncertain conditions.

Search process for living active systems takes place in a complex crowded environment. Our effort has been to capture the effect of simple confinement in a simple resetting mediated search process. However, we posit that these ideas could be extremely relevant to sub-cellular environment where there can be more obstacles such as macro-molecules which can hinder the search process. Also there can be chemical reactions between the constituents, thus one needs to gauge the interaction appropriately to the microscopic dynamics  \cite{volpe2017topography,van2022role}. More generally, it would be an outstanding challenge to design and explore resetting mediated strategies in complex systems.

\section*{Acknowledgements}
C.H. is supported by DST-INSPIRE-Faculty grant (Grant No.
IFA17-PH193). A.P. acknowledges support from the Department of Atomic Energy, India.

\section*{Data availability}
The data that supports the findings of this study is available within the article. The MATLAB codes that have been used to do the simulations are openly available in the GitHub repository
\onlinecite{sar2022}.

%\begin{thebibliography}{99}

\appendix

\section{Effect of rotational diffusion constant ($D_R$) in the first passage time fluctuations}
\label{appendixa}

In the main text, we have presented results mostly for a fixed rotational diffusion constant ($D_R$). In this appendix section, we relax this condition and investigate the effects of $D_R$ on the various first passage properties. In particular, we modulate $D_R$ and first study the variations of mean, fluctuations of the underlying first passage times. We simulate the underlying ABP dynamics (in the absence of resetting) and collect the first passage times from which we compute these observables for each $D_R$. Our first observation is that mean search time increases as a function of the distance between the target and the origin for any value of $D_R$ (irrespective of the initial conditions). The mean first passage time for a particular target also increases when the value of $D_R$ is increased.

\begin{figure}[b]
	\centerline{
		\includegraphics[scale = 0.033]{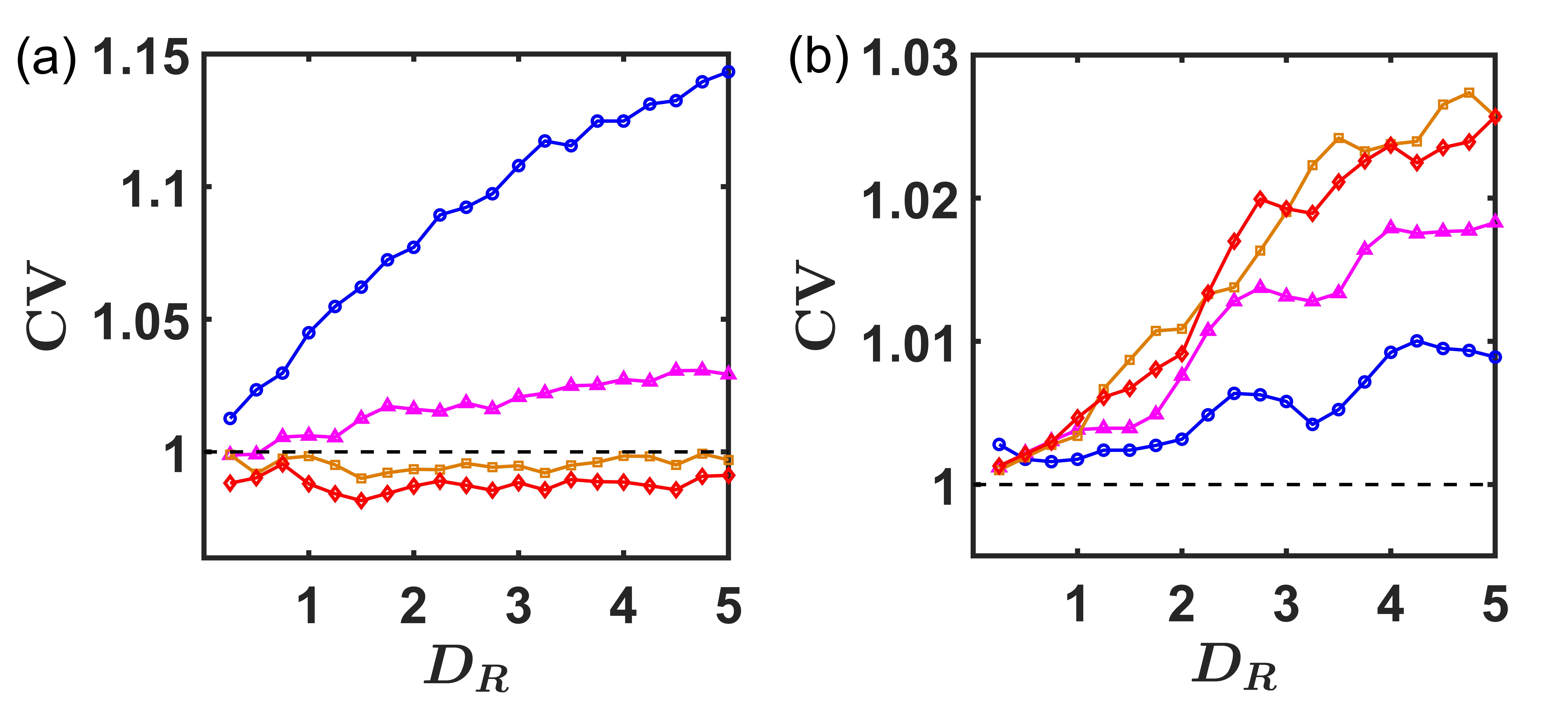}}
	\caption{{\bf  Modulation of the coefficient of variation ($\rm CV$) as a function of diffusion constant ($D_R$)}. $\rm CV$ of the underlying (\textit{without resetting}) first passage times as a function of rotational diffusion constant ($D_R$) is shown here for the four different targets. The separatrix $\rm CV = 1$ line is represented by the black dashed line. Panel (a): With origin as the source, the $\rm CV$ of the first passage times of targets $\rm A$ (blue) and $\rm B$ (magenta) increases from 1 as $D_R$ increases. However, the $\rm CV$ of the first passage times for targets $\rm C$ (yellow) and $\rm D$ (red) is always less than the $\rm CV = 1$ line. Panel (b): For random initial conditions, the $\rm CV$ values for all four targets are found to be greater than unity.}

	\label{fig7}
\end{figure}

Next, we would like to study the fluctuations as a function of $D_R$. To this end, we plot $\rm CV$ curves for different targets $\rm A$, $\rm B$, $\rm C$, and $\rm D$
(blue, magenta, yellow, and red, respectively) as a function of $D_R$ in Fig.\ \ref{fig7}. When the initial condition is fixed to the origin (IC1), we plot the variation of 
$\rm CV$
for all the targets by increasing the value of 
$D_R$ 
in Fig.\ \ref{fig7}(a). For the targets 
$\rm A$ and $\rm B$, 
the value of $\rm CV$ 
is greater than $1$ for almost all values of $D_R$. 
Contrarily, the value of 
$\rm CV$ 
is always less than
 $1$ for $\rm C$ as well as for $\rm D$. 
In Fig.\ \ref{fig7}(b), we present the variation of 
$\rm CV$ 
with increasing 
$D_R$ 
when the initial condition is chosen at random in every trial (IC2). It is observed that for $D_R \ge 1$, 
$\rm CV$ of all the targets is greater than $1$, i.e., resetting will certainly work if $D_R \ge 1$.
For $D_R<1$, the $\rm CV$ values are approximately close to 1. We accumulate the first passage time statistics in Table \ref{table2} for three different values of $D_R$.

\begin{table}[t]
	\centering
	\caption{Numerically calculated values of mean first passage time ($\langle T \rangle$), standard deviation ($\sigma$), and coefficient of variation ($\rm CV$) for four targets $\rm A$, $\rm B$, $\rm C$, and $\rm D$.  Here we take three different values of $D_R$ (viz. $D_R = 0.5$, $D_R  = 2.0$, and $D_R = 4.0$).}
	
	\begin{tabular}{|c|c|p{1.02cm}|p{1.02cm}|p{1.02cm}||p{1.02cm}|p{1.02cm}|p{1.02cm}|}
		\hline $D_R$ & Target & \multicolumn{3}{c||} {IC1} & \multicolumn{3}{c|} {IC2} \\
		\cline { 3 - 8 } & &$\langle T \rangle$ & $\sigma$ & $\rm CV$ & $\langle T \rangle$ & $\sigma$ & $\rm CV$ \\ \hline
		\cline { 1 - 8 } & $\rm A$ & 487.62 & 499.04 & 1.0234 & 498.33 & 499.28 & 1.0019 \\
		\cline { 2 - 8 } & $\rm B$ & 506.00 & 506.53 & 1.0010 & 506.98 & 508.55 & 1.0031\\
		\cline { 2 - 8 } 0.5 & $\rm C$ & 522.90 & 518.60 & 0.9917 & 521.44 & 522.74 & 1.0025 \\
		\cline { 2 - 8 } & $\rm D$ & 555.52 & 550.09 & 0.9902 & 549.92 & 551.05 & 1.0021 \\
		\hline
		\cline { 1 - 8 } & $\rm A$ & 543.98 & 585.93 & 1.0771 & 585.76 & 587.73 & 1.0034 \\
		\cline { 2 - 8 } & $\rm B$ & 609.22 & 619.05 & 1.0161 & 614.69 & 619.86 & 1.0084\\
		\cline { 2 - 8 } 2.0 & $\rm C$ & 683.82 & 679.34 & 0.9934 & 673.50 & 680.56 & 1.0105 \\
		\cline { 2 - 8 } & $\rm D$ & 816.56 & 806.05 & 0.9871 & 792.83 & 801.81 & 1.0113 \\
		\hline
		\cline { 1 - 8 } & $\rm A$ & 676.35 & 760.70 & 1.1247 & 761.34 & 770.35 & 1.0118 \\
		\cline { 2 - 8 } & $\rm B$ & 807.56 & 829.63 & 1.0273 & 820.17 & 836.95 & 1.0204\\
		\cline { 2 - 8 } 4.0 & $\rm C$ & 959.91 & 958.43 & 0.9984 & 935.39 & 956.28 & 1.0223 \\
		\cline { 2 - 8 } & $\rm D$ & 1212.90 & 1199.10 & 0.9886 & 1178.90 & 1203.30 & 1.0207 \\
		\hline
		
	\end{tabular}\label{table2}
\end{table}

\begin{figure*}[hpt]
	\centerline{
		\includegraphics[scale = 0.043]{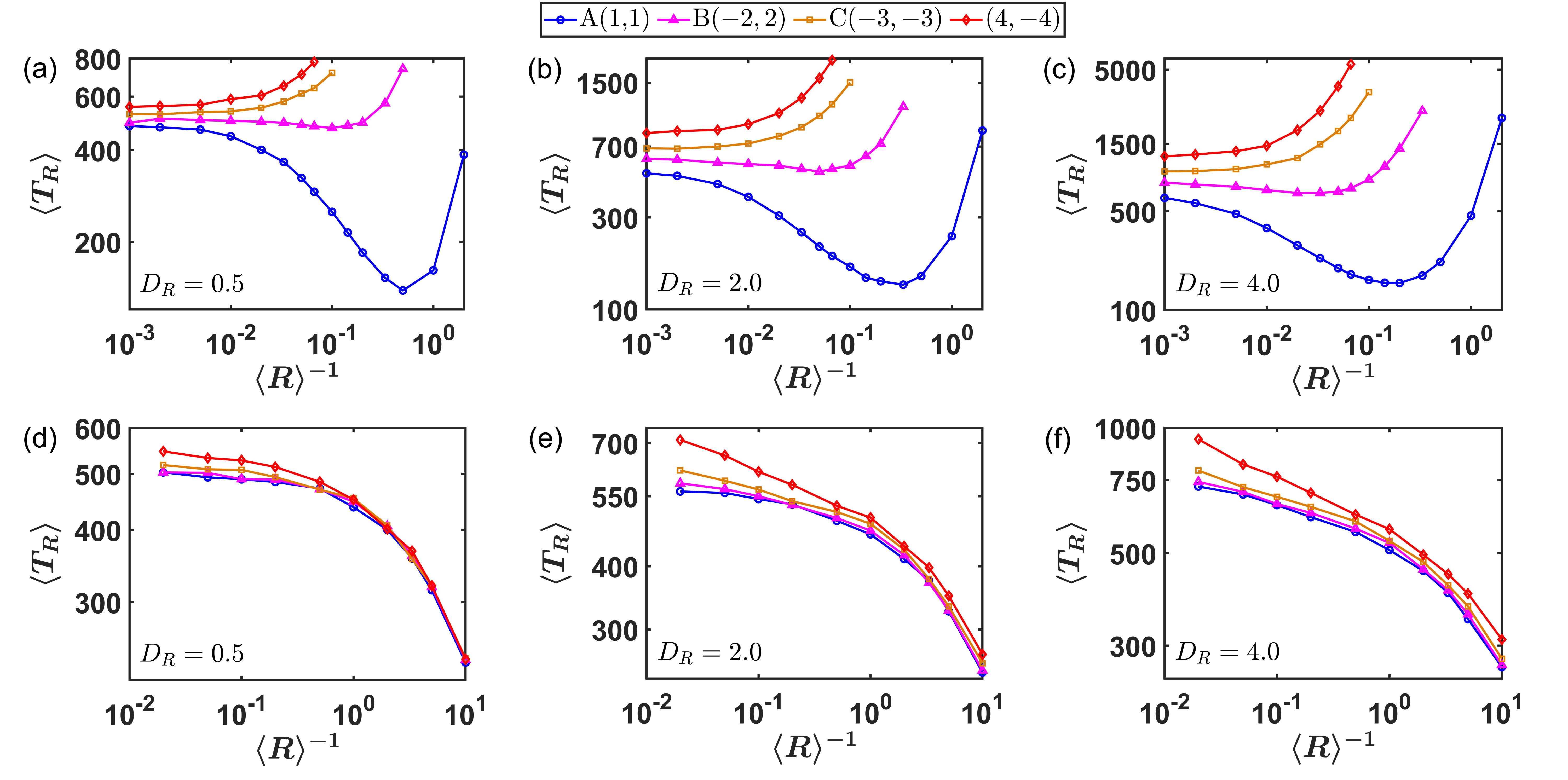}}
	\caption{{\bf Effect of rotational diffusion constant ($D_R$) on $\langle T_R \rangle$}. The top row contains plots designed for resetting protocol I, while the bottom row contains plots designed for resetting protocol II. For both the resetting protocols, we plot $\langle T_R \rangle$ as a function of resetting rate $\langle R^{-1} \rangle$. Three different $D_R$ values (0.5, 2.0, and 4.0) are selected. $D_R$ for the left hand panel ((a) and (d)) is 0.5, for middle panel ((b) and (e)) is 2.0, for the right-hand panel ((c) and (f)) is 4.0. For all three $D_R$ values, $\langle T_R \rangle$ for targets $\rm A$ (blue curve) and $\rm B$ (magenta curve) decrease under resetting protocol I. However, for the targets $\rm C$ (yellow curve) and $\rm D$ (red curve),  $\langle T_R \rangle$ always increases regardless of $D_R$ and $\langle R^{-1} \rangle$ (a direct ramification of $\rm CV <1$ -- also see Table\ \ref{table2}). For protocol II (bottom row), $\langle T_R \rangle$ is reduced as the resetting rate $\langle R \rangle^{-1}$ is gradually increased  (again, a direct ramification of the criterion $\rm CV >1$ -- also see Table\ \ref{table2}). This qualitative behaviour does not change as we increase $D_R$ further.}

	\label{dr-reset}
\end{figure*}

We now turn our attention to the first passage time statistics in the presence of resetting. 
The discussions from above (namely $\rm CV$ values of the underlying first passage times) provide us useful cues to predict the behavior of mean search time $\langle T_R \rangle$. The results are summarized in Fig.\ \ref{dr-reset} where we show the effect of resetting on the mean search times for different $D_R$. In the upper panel of Fig.\ \ref{dr-reset}, we present the results for resetting protocol I. It can be seen that $\langle T_R \rangle$ for target $\rm A$ decreases significantly for all three $D_R$ values which is clearly in accordance with the $\rm CV$ criterion (Eq.\ \eqref{cv}) for IC1 (also see Table \ref{table2}). The average search time is only marginally reduced for target $\rm B$. For other two targets $\rm C$ and $\rm D$, MFPT always increases which should not be surprising since $\rm CV$ always falls below 1 in this case. Now we move to resetting protocol II (or IC2) which is illustrated in the lower panel of Fig.\ \ref{dr-reset}. Here resetting successfully achieves to expedite the search process reducing $\langle T_R \rangle$ drastically for all $D_R$. This observation is indeed consistent with the numerical values obtained for $\rm CV$ (see Table \ref{table2}) which is always above unity for any values of $D_R$. An enhancement of $D_R$ results in large first passage time fluctuations which essentially add to the inherent fluctuations present in the system due to the randomness in the resetting/initial conditions (for IC2). These two sources of randomness essentially play a key role in the reduction of the overall search time under reset.

\section{Spatial spread of ABP under resetting}
\label{appendixb}
\begin{figure*}[hpt]
	\centerline{
		\includegraphics[scale = 0.056]{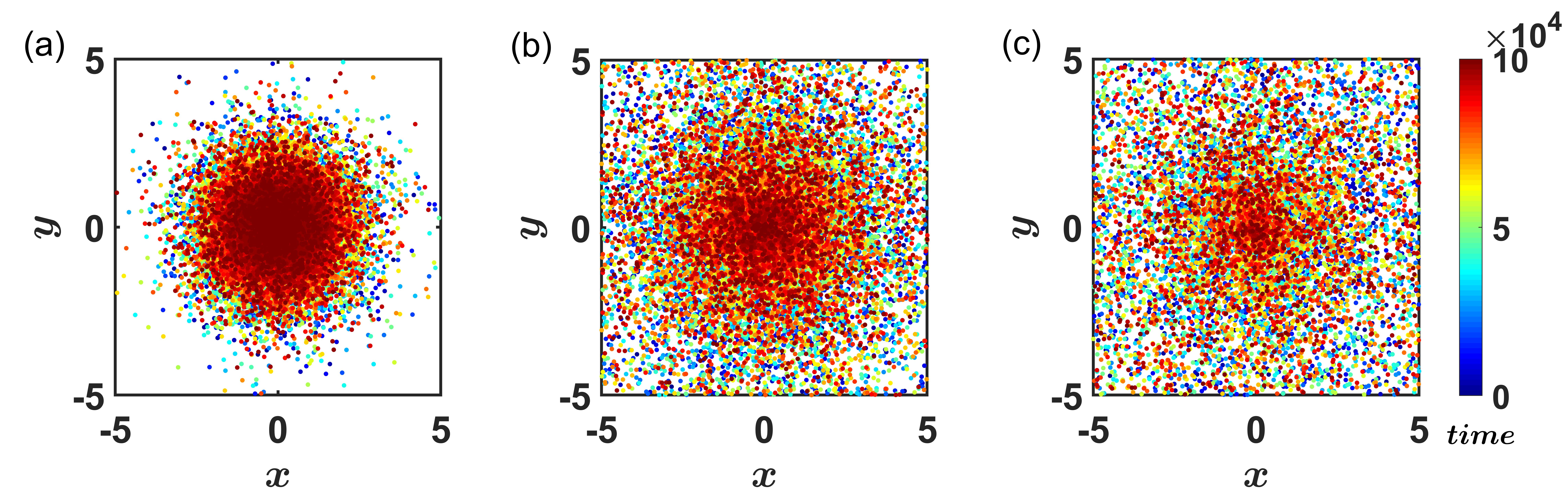}}
	\caption{{\bf Spatial spread of the ABP in the phase space spanned over $(x,y)$ plane under resetting protocol I.} In the numerical simulation, at each microscopic time step we have measured position $(x,y)$ of the ABP and have marked the same as a single point on the phase space. Total time step taken is $10^5$ units. This clearly indicates that the particle is most probable to be found near the origin for all resetting times, but the spread keeps growing as we increase the mean resetting time. Resetting times were drawn from the distribution $1/\langle R \rangle e^{-t/\langle R \rangle}$ with the following values for $\langle R \rangle$:  $1.0$ (panel a), $5.0$ (panel b), and $10.0$ (panel c).}
	\label{spread1}
\end{figure*}

\begin{figure}[hpt]
	\centerline{
		\includegraphics[scale = 0.07]{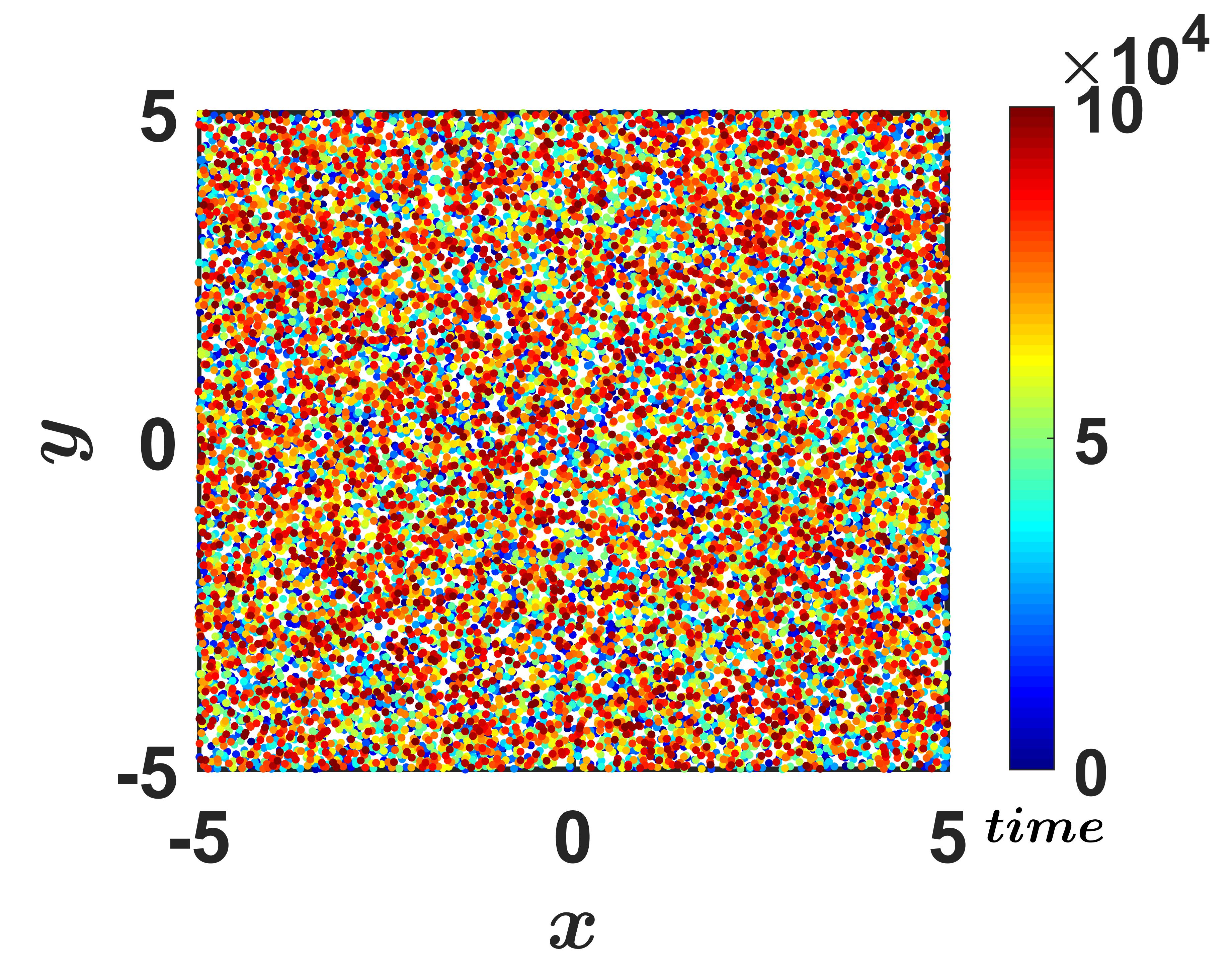}}
	\caption{{\bf Spatial spread of the ABP in the phase space under resetting protocol II.} In the numerical simulation, at each microscopic time step we have measured position $(x,y)$ of the ABP and have marked the same as a single point on the phase space. Total time step taken is $10^5$ units. Looking at the homogeneous spread of the phase space points, it is evident that the particle equally likely to be found anywhere in the phase space. Resetting times were drawn from the distribution $1/\langle R \rangle e^{-t/\langle R \rangle}$ with $\langle R \rangle =5.0$.} 
	\label{spread2}
\end{figure}

In this section, we delineate the phase space of the ABP under both the resetting protocols. Essentially, here, we have just taken a single snapshot of the ABP's position $(x,y)$ at a given time step up to $10^5$ time units (see Figs. \ref{spread1} and \ref{spread2}). In both cases, each dot represents the position of ABP in the 2D plane and these points are colored according to the observation time scale. Since we have not imposed any target here, the particle keeps evolving and the distribution is expected to reach a stationary state. Taking such data set for many realizations, we have performed an ensemble average which resulted in the probability distribution function. As can be seen from Fig. \ref{spread1}, the ABP remains close to the origin since it is reset to the origin. This indicates that the stationary state will be peaked around the origin with highest probability as shown in Fig. \ref{fig4}. Similarly, for the randomized resetting condition, the positions are spread all over the phase space as it is shown in Fig. \ref{spread2}. This renders a probability distribution function which is uniform over the topography as is evident from Fig. \ref{fig10}.

\nocite{*}	
\bibliography{resetting}

\providecommand{\noopsort}[1]{}\providecommand{\singleletter}[1]{#1}%
\begin{thebibliography}{77}%
\makeatletter
\providecommand \@ifxundefined [1]{%
 \@ifx{#1\undefined}
}%
\providecommand \@ifnum [1]{%
 \ifnum #1\expandafter \@firstoftwo
 \else \expandafter \@secondoftwo
 \fi
}%
\providecommand \@ifx [1]{%
 \ifx #1\expandafter \@firstoftwo
 \else \expandafter \@secondoftwo
 \fi
}%
\providecommand \natexlab [1]{#1}%
\providecommand \enquote  [1]{``#1''}%
\providecommand \bibnamefont  [1]{#1}%
\providecommand \bibfnamefont [1]{#1}%
\providecommand \citenamefont [1]{#1}%
\providecommand \href@noop [0]{\@secondoftwo}%
\providecommand \href [0]{\begingroup \@sanitize@url \@href}%
\providecommand \@href[1]{\@@startlink{#1}\@@href}%
\providecommand \@@href[1]{\endgroup#1\@@endlink}%
\providecommand \@sanitize@url [0]{\catcode `\\12\catcode `\$12\catcode
  `\&12\catcode `\#12\catcode `\^12\catcode `\_12\catcode `\%12\relax}%
\providecommand \@@startlink[1]{}%
\providecommand \@@endlink[0]{}%
\providecommand \url  [0]{\begingroup\@sanitize@url \@url }%
\providecommand \@url [1]{\endgroup\@href {#1}{\urlprefix }}%
\providecommand \urlprefix  [0]{URL }%
\providecommand \Eprint [0]{\href }%
\providecommand \doibase [0]{http://dx.doi.org/}%
\providecommand \selectlanguage [0]{\@gobble}%
\providecommand \bibinfo  [0]{\@secondoftwo}%
\providecommand \bibfield  [0]{\@secondoftwo}%
\providecommand \translation [1]{[#1]}%
\providecommand \BibitemOpen [0]{}%
\providecommand \bibitemStop [0]{}%
\providecommand \bibitemNoStop [0]{.\EOS\space}%
\providecommand \EOS [0]{\spacefactor3000\relax}%
\providecommand \BibitemShut  [1]{\csname bibitem#1\endcsname}%
\let\auto@bib@innerbib\@empty
%</preamble>
\bibitem [{\citenamefont {Toner}, \citenamefont {Tu},\ and\ \citenamefont
  {Ramaswamy}(2005)}]{toner2005hydrodynamics}%
  \BibitemOpen
  \bibfield  {author} {\bibinfo {author} {\bibfnamefont {J.}~\bibnamefont
  {Toner}}, \bibinfo {author} {\bibfnamefont {Y.}~\bibnamefont {Tu}}, \ and\
  \bibinfo {author} {\bibfnamefont {S.}~\bibnamefont {Ramaswamy}},\ }\bibfield
  {title} {\enquote {\bibinfo {title} {Hydrodynamics and phases of flocks},}\
  }\href@noop {} {\bibfield  {journal} {\bibinfo  {journal} {Annals of
  Physics}\ }\textbf {\bibinfo {volume} {318}},\ \bibinfo {pages} {170--244}
  (\bibinfo {year} {2005})}\BibitemShut {NoStop}%
\bibitem [{\citenamefont {Ramaswamy}(2010)}]{SR-review}%
  \BibitemOpen
  \bibfield  {author} {\bibinfo {author} {\bibfnamefont {S.}~\bibnamefont
  {Ramaswamy}},\ }\bibfield  {title} {\enquote {\bibinfo {title} {The mechanics
  and statistics of active matter},}\ }\href@noop {} {\bibfield  {journal}
  {\bibinfo  {journal} {Annual Review of Condensed Matter Physics}\ }\textbf
  {\bibinfo {volume} {1}},\ \bibinfo {pages} {323--345} (\bibinfo {year}
  {2010})}\BibitemShut {NoStop}%
\bibitem [{\citenamefont {Romanczuk}\ \emph {et~al.}(2012)\citenamefont
  {Romanczuk}, \citenamefont {B{\"a}r}, \citenamefont {Ebeling}, \citenamefont
  {Lindner},\ and\ \citenamefont {Schimansky-Geier}}]{romanczuk2012active}%
  \BibitemOpen
  \bibfield  {author} {\bibinfo {author} {\bibfnamefont {P.}~\bibnamefont
  {Romanczuk}}, \bibinfo {author} {\bibfnamefont {M.}~\bibnamefont {B{\"a}r}},
  \bibinfo {author} {\bibfnamefont {W.}~\bibnamefont {Ebeling}}, \bibinfo
  {author} {\bibfnamefont {B.}~\bibnamefont {Lindner}}, \ and\ \bibinfo
  {author} {\bibfnamefont {L.}~\bibnamefont {Schimansky-Geier}},\ }\bibfield
  {title} {\enquote {\bibinfo {title} {Active brownian particles},}\
  }\href@noop {} {\bibfield  {journal} {\bibinfo  {journal} {The European
  Physical Journal Special Topics}\ }\textbf {\bibinfo {volume} {202}},\
  \bibinfo {pages} {1--162} (\bibinfo {year} {2012})}\BibitemShut {NoStop}%
\bibitem [{\citenamefont {Gnesotto}\ \emph {et~al.}(2018)\citenamefont
  {Gnesotto}, \citenamefont {Mura}, \citenamefont {Gladrow},\ and\
  \citenamefont {Broedersz}}]{gnesotto2018broken}%
  \BibitemOpen
  \bibfield  {author} {\bibinfo {author} {\bibfnamefont {F.~S.}\ \bibnamefont
  {Gnesotto}}, \bibinfo {author} {\bibfnamefont {F.}~\bibnamefont {Mura}},
  \bibinfo {author} {\bibfnamefont {J.}~\bibnamefont {Gladrow}}, \ and\
  \bibinfo {author} {\bibfnamefont {C.~P.}\ \bibnamefont {Broedersz}},\
  }\bibfield  {title} {\enquote {\bibinfo {title} {Broken detailed balance and
  non-equilibrium dynamics in living systems: a review},}\ }\href@noop {}
  {\bibfield  {journal} {\bibinfo  {journal} {Reports on Progress in Physics}\
  }\textbf {\bibinfo {volume} {81}},\ \bibinfo {pages} {066601} (\bibinfo
  {year} {2018})}\BibitemShut {NoStop}%
\bibitem [{\citenamefont {Berg}(2008)}]{berg2008coli}%
  \BibitemOpen
  \bibfield  {author} {\bibinfo {author} {\bibfnamefont {H.~C.}\ \bibnamefont
  {Berg}},\ }\href@noop {} {\emph {\bibinfo {title} {E. coli in Motion}}}\
  (\bibinfo  {publisher} {Springer Science \& Business Media},\ \bibinfo {year}
  {2008})\BibitemShut {NoStop}%
\bibitem [{\citenamefont {Berg}(2018)}]{berg2018random}%
  \BibitemOpen
  \bibfield  {author} {\bibinfo {author} {\bibfnamefont {H.~C.}\ \bibnamefont
  {Berg}},\ }\href@noop {} {\emph {\bibinfo {title} {Random walks in
  biology}}}\ (\bibinfo  {publisher} {Princeton University Press},\ \bibinfo
  {year} {2018})\BibitemShut {NoStop}%
\bibitem [{\citenamefont {Jiang}, \citenamefont {Yoshinaga},\ and\
  \citenamefont {Sano}(2010)}]{jiang2010active}%
  \BibitemOpen
  \bibfield  {author} {\bibinfo {author} {\bibfnamefont {H.-R.}\ \bibnamefont
  {Jiang}}, \bibinfo {author} {\bibfnamefont {N.}~\bibnamefont {Yoshinaga}}, \
  and\ \bibinfo {author} {\bibfnamefont {M.}~\bibnamefont {Sano}},\ }\bibfield
  {title} {\enquote {\bibinfo {title} {Active motion of a janus particle by
  self-thermophoresis in a defocused laser beam},}\ }\href@noop {} {\bibfield
  {journal} {\bibinfo  {journal} {Physical Review Letters}\ }\textbf {\bibinfo
  {volume} {105}},\ \bibinfo {pages} {268302} (\bibinfo {year}
  {2010})}\BibitemShut {NoStop}%
\bibitem [{\citenamefont {Mart{\'\i}n-G{\'o}mez}\ \emph
  {et~al.}(2018)\citenamefont {Mart{\'\i}n-G{\'o}mez}, \citenamefont {Levis},
  \citenamefont {D{\'\i}az-Guilera},\ and\ \citenamefont
  {Pagonabarraga}}]{martin2018collective}%
  \BibitemOpen
  \bibfield  {author} {\bibinfo {author} {\bibfnamefont {A.}~\bibnamefont
  {Mart{\'\i}n-G{\'o}mez}}, \bibinfo {author} {\bibfnamefont {D.}~\bibnamefont
  {Levis}}, \bibinfo {author} {\bibfnamefont {A.}~\bibnamefont
  {D{\'\i}az-Guilera}}, \ and\ \bibinfo {author} {\bibfnamefont
  {I.}~\bibnamefont {Pagonabarraga}},\ }\bibfield  {title} {\enquote {\bibinfo
  {title} {Collective motion of active brownian particles with polar
  alignment},}\ }\href@noop {} {\bibfield  {journal} {\bibinfo  {journal} {Soft
  matter}\ }\textbf {\bibinfo {volume} {14}},\ \bibinfo {pages} {2610--2618}
  (\bibinfo {year} {2018})}\BibitemShut {NoStop}%
\bibitem [{\citenamefont {Gompper}\ \emph {et~al.}(2020)\citenamefont
  {Gompper}, \citenamefont {Winkler}, \citenamefont {Speck}, \citenamefont
  {Solon}, \citenamefont {Nardini}, \citenamefont {Peruani}, \citenamefont
  {L{\"o}wen}, \citenamefont {Golestanian}, \citenamefont {Kaupp},
  \citenamefont {Alvarez} \emph {et~al.}}]{gompper20202020}%
  \BibitemOpen
  \bibfield  {author} {\bibinfo {author} {\bibfnamefont {G.}~\bibnamefont
  {Gompper}}, \bibinfo {author} {\bibfnamefont {R.~G.}\ \bibnamefont
  {Winkler}}, \bibinfo {author} {\bibfnamefont {T.}~\bibnamefont {Speck}},
  \bibinfo {author} {\bibfnamefont {A.}~\bibnamefont {Solon}}, \bibinfo
  {author} {\bibfnamefont {C.}~\bibnamefont {Nardini}}, \bibinfo {author}
  {\bibfnamefont {F.}~\bibnamefont {Peruani}}, \bibinfo {author} {\bibfnamefont
  {H.}~\bibnamefont {L{\"o}wen}}, \bibinfo {author} {\bibfnamefont
  {R.}~\bibnamefont {Golestanian}}, \bibinfo {author} {\bibfnamefont {U.~B.}\
  \bibnamefont {Kaupp}}, \bibinfo {author} {\bibfnamefont {L.}~\bibnamefont
  {Alvarez}},  \emph {et~al.},\ }\bibfield  {title} {\enquote {\bibinfo {title}
  {The 2020 motile active matter roadmap},}\ }\href@noop {} {\bibfield
  {journal} {\bibinfo  {journal} {Journal of Physics: Condensed Matter}\
  }\textbf {\bibinfo {volume} {32}},\ \bibinfo {pages} {193001} (\bibinfo
  {year} {2020})}\BibitemShut {NoStop}%
\bibitem [{\citenamefont {Mart{\'\i}nez-Calvo}, \citenamefont {Trenado-Yuste},\
  and\ \citenamefont {Datta}(2021)}]{martinez2021active}%
  \BibitemOpen
  \bibfield  {author} {\bibinfo {author} {\bibfnamefont {A.}~\bibnamefont
  {Mart{\'\i}nez-Calvo}}, \bibinfo {author} {\bibfnamefont {C.}~\bibnamefont
  {Trenado-Yuste}}, \ and\ \bibinfo {author} {\bibfnamefont {S.~S.}\
  \bibnamefont {Datta}},\ }\bibfield  {title} {\enquote {\bibinfo {title}
  {Active transport in complex environments},}\ }\href@noop {} {\bibfield
  {journal} {\bibinfo  {journal} {arXiv preprint arXiv:2108.07011}\ } (\bibinfo
  {year} {2021})}\BibitemShut {NoStop}%
\bibitem [{\citenamefont {Mijalkov}\ and\ \citenamefont
  {Volpe}(2013)}]{mijalkov2013sorting}%
  \BibitemOpen
  \bibfield  {author} {\bibinfo {author} {\bibfnamefont {M.}~\bibnamefont
  {Mijalkov}}\ and\ \bibinfo {author} {\bibfnamefont {G.}~\bibnamefont
  {Volpe}},\ }\bibfield  {title} {\enquote {\bibinfo {title} {Sorting of chiral
  microswimmers},}\ }\href@noop {} {\bibfield  {journal} {\bibinfo  {journal}
  {Soft Matter}\ }\textbf {\bibinfo {volume} {9}},\ \bibinfo {pages}
  {6376--6381} (\bibinfo {year} {2013})}\BibitemShut {NoStop}%
\bibitem [{\citenamefont {Bechinger}\ \emph {et~al.}(2016)\citenamefont
  {Bechinger}, \citenamefont {Di~Leonardo}, \citenamefont {L{\"o}wen},
  \citenamefont {Reichhardt}, \citenamefont {Volpe},\ and\ \citenamefont
  {Volpe}}]{bechinger2016active}%
  \BibitemOpen
  \bibfield  {author} {\bibinfo {author} {\bibfnamefont {C.}~\bibnamefont
  {Bechinger}}, \bibinfo {author} {\bibfnamefont {R.}~\bibnamefont
  {Di~Leonardo}}, \bibinfo {author} {\bibfnamefont {H.}~\bibnamefont
  {L{\"o}wen}}, \bibinfo {author} {\bibfnamefont {C.}~\bibnamefont
  {Reichhardt}}, \bibinfo {author} {\bibfnamefont {G.}~\bibnamefont {Volpe}}, \
  and\ \bibinfo {author} {\bibfnamefont {G.}~\bibnamefont {Volpe}},\ }\bibfield
   {title} {\enquote {\bibinfo {title} {Active particles in complex and crowded
  environments},}\ }\href@noop {} {\bibfield  {journal} {\bibinfo  {journal}
  {Reviews of Modern Physics}\ }\textbf {\bibinfo {volume} {88}},\ \bibinfo
  {pages} {045006} (\bibinfo {year} {2016})}\BibitemShut {NoStop}%
\bibitem [{\citenamefont {Volpe}\ \emph {et~al.}(2011)\citenamefont {Volpe},
  \citenamefont {Buttinoni}, \citenamefont {Vogt}, \citenamefont
  {K{\"u}mmerer},\ and\ \citenamefont {Bechinger}}]{volpe2011microswimmers}%
  \BibitemOpen
  \bibfield  {author} {\bibinfo {author} {\bibfnamefont {G.}~\bibnamefont
  {Volpe}}, \bibinfo {author} {\bibfnamefont {I.}~\bibnamefont {Buttinoni}},
  \bibinfo {author} {\bibfnamefont {D.}~\bibnamefont {Vogt}}, \bibinfo {author}
  {\bibfnamefont {H.-J.}\ \bibnamefont {K{\"u}mmerer}}, \ and\ \bibinfo
  {author} {\bibfnamefont {C.}~\bibnamefont {Bechinger}},\ }\bibfield  {title}
  {\enquote {\bibinfo {title} {Microswimmers in patterned environments},}\
  }\href@noop {} {\bibfield  {journal} {\bibinfo  {journal} {Soft Matter}\
  }\textbf {\bibinfo {volume} {7}},\ \bibinfo {pages} {8810--8815} (\bibinfo
  {year} {2011})}\BibitemShut {NoStop}%
\bibitem [{\citenamefont {Ross}, \citenamefont {Ali},\ and\ \citenamefont
  {Warshaw}(2008)}]{ross2008cargo}%
  \BibitemOpen
  \bibfield  {author} {\bibinfo {author} {\bibfnamefont {J.~L.}\ \bibnamefont
  {Ross}}, \bibinfo {author} {\bibfnamefont {M.~Y.}\ \bibnamefont {Ali}}, \
  and\ \bibinfo {author} {\bibfnamefont {D.~M.}\ \bibnamefont {Warshaw}},\
  }\bibfield  {title} {\enquote {\bibinfo {title} {Cargo transport: molecular
  motors navigate a complex cytoskeleton},}\ }\href@noop {} {\bibfield
  {journal} {\bibinfo  {journal} {Current opinion in cell biology}\ }\textbf
  {\bibinfo {volume} {20}},\ \bibinfo {pages} {41--47} (\bibinfo {year}
  {2008})}\BibitemShut {NoStop}%
\bibitem [{\citenamefont {Ghosh}\ \emph {et~al.}(2020)\citenamefont {Ghosh},
  \citenamefont {Xu}, \citenamefont {Gupta},\ and\ \citenamefont
  {Gracias}}]{ghosh2020active}%
  \BibitemOpen
  \bibfield  {author} {\bibinfo {author} {\bibfnamefont {A.}~\bibnamefont
  {Ghosh}}, \bibinfo {author} {\bibfnamefont {W.}~\bibnamefont {Xu}}, \bibinfo
  {author} {\bibfnamefont {N.}~\bibnamefont {Gupta}}, \ and\ \bibinfo {author}
  {\bibfnamefont {D.~H.}\ \bibnamefont {Gracias}},\ }\bibfield  {title}
  {\enquote {\bibinfo {title} {Active matter therapeutics},}\ }\href@noop {}
  {\bibfield  {journal} {\bibinfo  {journal} {Nano Today}\ }\textbf {\bibinfo
  {volume} {31}},\ \bibinfo {pages} {100836} (\bibinfo {year}
  {2020})}\BibitemShut {NoStop}%
\bibitem [{\citenamefont {B{\'e}nichou}\ \emph {et~al.}(2011)\citenamefont
  {B{\'e}nichou}, \citenamefont {Loverdo}, \citenamefont {Moreau},\ and\
  \citenamefont {Voituriez}}]{benichou2011intermittent}%
  \BibitemOpen
  \bibfield  {author} {\bibinfo {author} {\bibfnamefont {O.}~\bibnamefont
  {B{\'e}nichou}}, \bibinfo {author} {\bibfnamefont {C.}~\bibnamefont
  {Loverdo}}, \bibinfo {author} {\bibfnamefont {M.}~\bibnamefont {Moreau}}, \
  and\ \bibinfo {author} {\bibfnamefont {R.}~\bibnamefont {Voituriez}},\
  }\bibfield  {title} {\enquote {\bibinfo {title} {Intermittent search
  strategies},}\ }\href@noop {} {\bibfield  {journal} {\bibinfo  {journal}
  {Reviews of Modern Physics}\ }\textbf {\bibinfo {volume} {83}},\ \bibinfo
  {pages} {81} (\bibinfo {year} {2011})}\BibitemShut {NoStop}%
\bibitem [{\citenamefont {Wang}\ \emph {et~al.}(2017)\citenamefont {Wang},
  \citenamefont {Zhang}, \citenamefont {Xia},\ and\ \citenamefont
  {Yu}}]{wang2017spatial}%
  \BibitemOpen
  \bibfield  {author} {\bibinfo {author} {\bibfnamefont {J.}~\bibnamefont
  {Wang}}, \bibinfo {author} {\bibfnamefont {D.}~\bibnamefont {Zhang}},
  \bibinfo {author} {\bibfnamefont {B.}~\bibnamefont {Xia}}, \ and\ \bibinfo
  {author} {\bibfnamefont {W.}~\bibnamefont {Yu}},\ }\bibfield  {title}
  {\enquote {\bibinfo {title} {Spatial heterogeneity can facilitate the target
  search of self-propelled particles},}\ }\href@noop {} {\bibfield  {journal}
  {\bibinfo  {journal} {Soft Matter}\ }\textbf {\bibinfo {volume} {13}},\
  \bibinfo {pages} {758--764} (\bibinfo {year} {2017})}\BibitemShut {NoStop}%
\bibitem [{\citenamefont {Malakar}\ \emph {et~al.}(2018)\citenamefont
  {Malakar}, \citenamefont {Jemseena}, \citenamefont {Kundu}, \citenamefont
  {Kumar}, \citenamefont {Sabhapandit}, \citenamefont {Majumdar}, \citenamefont
  {Redner},\ and\ \citenamefont {Dhar}}]{malakar2018steady}%
  \BibitemOpen
  \bibfield  {author} {\bibinfo {author} {\bibfnamefont {K.}~\bibnamefont
  {Malakar}}, \bibinfo {author} {\bibfnamefont {V.}~\bibnamefont {Jemseena}},
  \bibinfo {author} {\bibfnamefont {A.}~\bibnamefont {Kundu}}, \bibinfo
  {author} {\bibfnamefont {K.~V.}\ \bibnamefont {Kumar}}, \bibinfo {author}
  {\bibfnamefont {S.}~\bibnamefont {Sabhapandit}}, \bibinfo {author}
  {\bibfnamefont {S.~N.}\ \bibnamefont {Majumdar}}, \bibinfo {author}
  {\bibfnamefont {S.}~\bibnamefont {Redner}}, \ and\ \bibinfo {author}
  {\bibfnamefont {A.}~\bibnamefont {Dhar}},\ }\bibfield  {title} {\enquote
  {\bibinfo {title} {Steady state, relaxation and first-passage properties of a
  run-and-tumble particle in one-dimension},}\ }\href@noop {} {\bibfield
  {journal} {\bibinfo  {journal} {Journal of Statistical Mechanics: Theory and
  Experiment}\ }\textbf {\bibinfo {volume} {2018}},\ \bibinfo {pages} {043215}
  (\bibinfo {year} {2018})}\BibitemShut {NoStop}%
\bibitem [{\citenamefont {Angelani}, \citenamefont {Di~Leonardo},\ and\
  \citenamefont {Paoluzzi}(2014)}]{angelani2014first}%
  \BibitemOpen
  \bibfield  {author} {\bibinfo {author} {\bibfnamefont {L.}~\bibnamefont
  {Angelani}}, \bibinfo {author} {\bibfnamefont {R.}~\bibnamefont
  {Di~Leonardo}}, \ and\ \bibinfo {author} {\bibfnamefont {M.}~\bibnamefont
  {Paoluzzi}},\ }\bibfield  {title} {\enquote {\bibinfo {title} {First-passage
  time of run-and-tumble particles},}\ }\href@noop {} {\bibfield  {journal}
  {\bibinfo  {journal} {The European Physical Journal E}\ }\textbf {\bibinfo
  {volume} {37}},\ \bibinfo {pages} {1--6} (\bibinfo {year}
  {2014})}\BibitemShut {NoStop}%
\bibitem [{\citenamefont {Condamin}\ \emph {et~al.}(2007)\citenamefont
  {Condamin}, \citenamefont {B{\'e}nichou}, \citenamefont {Tejedor},
  \citenamefont {Voituriez},\ and\ \citenamefont
  {Klafter}}]{condamin2007first}%
  \BibitemOpen
  \bibfield  {author} {\bibinfo {author} {\bibfnamefont {S.}~\bibnamefont
  {Condamin}}, \bibinfo {author} {\bibfnamefont {O.}~\bibnamefont
  {B{\'e}nichou}}, \bibinfo {author} {\bibfnamefont {V.}~\bibnamefont
  {Tejedor}}, \bibinfo {author} {\bibfnamefont {R.}~\bibnamefont {Voituriez}},
  \ and\ \bibinfo {author} {\bibfnamefont {J.}~\bibnamefont {Klafter}},\
  }\bibfield  {title} {\enquote {\bibinfo {title} {First-passage times in
  complex scale-invariant media},}\ }\href@noop {} {\bibfield  {journal}
  {\bibinfo  {journal} {Nature}\ }\textbf {\bibinfo {volume} {450}},\ \bibinfo
  {pages} {77--80} (\bibinfo {year} {2007})}\BibitemShut {NoStop}%
\bibitem [{\citenamefont {B{\'e}nichou}\ \emph {et~al.}(2005)\citenamefont
  {B{\'e}nichou}, \citenamefont {Coppey}, \citenamefont {Moreau}, \citenamefont
  {Suet},\ and\ \citenamefont {Voituriez}}]{benichou2005optimal}%
  \BibitemOpen
  \bibfield  {author} {\bibinfo {author} {\bibfnamefont {O.}~\bibnamefont
  {B{\'e}nichou}}, \bibinfo {author} {\bibfnamefont {M.}~\bibnamefont
  {Coppey}}, \bibinfo {author} {\bibfnamefont {M.}~\bibnamefont {Moreau}},
  \bibinfo {author} {\bibfnamefont {P.}~\bibnamefont {Suet}}, \ and\ \bibinfo
  {author} {\bibfnamefont {R.}~\bibnamefont {Voituriez}},\ }\bibfield  {title}
  {\enquote {\bibinfo {title} {Optimal search strategies for hidden targets},}\
  }\href@noop {} {\bibfield  {journal} {\bibinfo  {journal} {Physical Review
  Letters}\ }\textbf {\bibinfo {volume} {94}},\ \bibinfo {pages} {198101}
  (\bibinfo {year} {2005})}\BibitemShut {NoStop}%
\bibitem [{\citenamefont {B{\'e}nichou}\ \emph {et~al.}(2010)\citenamefont
  {B{\'e}nichou}, \citenamefont {Chevalier}, \citenamefont {Klafter},
  \citenamefont {Meyer},\ and\ \citenamefont
  {Voituriez}}]{benichou2010geometry}%
  \BibitemOpen
  \bibfield  {author} {\bibinfo {author} {\bibfnamefont {O.}~\bibnamefont
  {B{\'e}nichou}}, \bibinfo {author} {\bibfnamefont {C.}~\bibnamefont
  {Chevalier}}, \bibinfo {author} {\bibfnamefont {J.}~\bibnamefont {Klafter}},
  \bibinfo {author} {\bibfnamefont {B.}~\bibnamefont {Meyer}}, \ and\ \bibinfo
  {author} {\bibfnamefont {R.}~\bibnamefont {Voituriez}},\ }\bibfield  {title}
  {\enquote {\bibinfo {title} {Geometry-controlled kinetics},}\ }\href@noop {}
  {\bibfield  {journal} {\bibinfo  {journal} {Nature Chemistry}\ }\textbf
  {\bibinfo {volume} {2}},\ \bibinfo {pages} {472--477} (\bibinfo {year}
  {2010})}\BibitemShut {NoStop}%
\bibitem [{\citenamefont {Bray}, \citenamefont {Majumdar},\ and\ \citenamefont
  {Schehr}(2013)}]{bray2013persistence}%
  \BibitemOpen
  \bibfield  {author} {\bibinfo {author} {\bibfnamefont {A.~J.}\ \bibnamefont
  {Bray}}, \bibinfo {author} {\bibfnamefont {S.~N.}\ \bibnamefont {Majumdar}},
  \ and\ \bibinfo {author} {\bibfnamefont {G.}~\bibnamefont {Schehr}},\
  }\bibfield  {title} {\enquote {\bibinfo {title} {Persistence and
  first-passage properties in nonequilibrium systems},}\ }\href@noop {}
  {\bibfield  {journal} {\bibinfo  {journal} {Advances in Physics}\ }\textbf
  {\bibinfo {volume} {62}},\ \bibinfo {pages} {225--361} (\bibinfo {year}
  {2013})}\BibitemShut {NoStop}%
\bibitem [{\citenamefont {Viswanathan}\ \emph {et~al.}(2011)\citenamefont
  {Viswanathan}, \citenamefont {Da~Luz}, \citenamefont {Raposo},\ and\
  \citenamefont {Stanley}}]{viswanathan2011physics}%
  \BibitemOpen
  \bibfield  {author} {\bibinfo {author} {\bibfnamefont {G.~M.}\ \bibnamefont
  {Viswanathan}}, \bibinfo {author} {\bibfnamefont {M.~G.}\ \bibnamefont
  {Da~Luz}}, \bibinfo {author} {\bibfnamefont {E.~P.}\ \bibnamefont {Raposo}},
  \ and\ \bibinfo {author} {\bibfnamefont {H.~E.}\ \bibnamefont {Stanley}},\
  }\href@noop {} {\emph {\bibinfo {title} {The physics of foraging: an
  introduction to random searches and biological encounters}}}\ (\bibinfo
  {publisher} {Cambridge University Press},\ \bibinfo {year}
  {2011})\BibitemShut {NoStop}%
\bibitem [{\citenamefont {Rhee}\ \emph {et~al.}(2011)\citenamefont {Rhee},
  \citenamefont {Shin}, \citenamefont {Hong}, \citenamefont {Lee},
  \citenamefont {Kim},\ and\ \citenamefont {Chong}}]{rhee2011levy}%
  \BibitemOpen
  \bibfield  {author} {\bibinfo {author} {\bibfnamefont {I.}~\bibnamefont
  {Rhee}}, \bibinfo {author} {\bibfnamefont {M.}~\bibnamefont {Shin}}, \bibinfo
  {author} {\bibfnamefont {S.}~\bibnamefont {Hong}}, \bibinfo {author}
  {\bibfnamefont {K.}~\bibnamefont {Lee}}, \bibinfo {author} {\bibfnamefont
  {S.~J.}\ \bibnamefont {Kim}}, \ and\ \bibinfo {author} {\bibfnamefont
  {S.}~\bibnamefont {Chong}},\ }\bibfield  {title} {\enquote {\bibinfo {title}
  {On the levy-walk nature of human mobility},}\ }\href@noop {} {\bibfield
  {journal} {\bibinfo  {journal} {IEEE/ACM Transactions on Networking}\
  }\textbf {\bibinfo {volume} {19}},\ \bibinfo {pages} {630--643} (\bibinfo
  {year} {2011})}\BibitemShut {NoStop}%
\bibitem [{\citenamefont {Zaburdaev}, \citenamefont {Denisov},\ and\
  \citenamefont {Klafter}(2015)}]{zaburdaev2015levy}%
  \BibitemOpen
  \bibfield  {author} {\bibinfo {author} {\bibfnamefont {V.}~\bibnamefont
  {Zaburdaev}}, \bibinfo {author} {\bibfnamefont {S.}~\bibnamefont {Denisov}},
  \ and\ \bibinfo {author} {\bibfnamefont {J.}~\bibnamefont {Klafter}},\
  }\bibfield  {title} {\enquote {\bibinfo {title} {L{\'e}vy walks},}\
  }\href@noop {} {\bibfield  {journal} {\bibinfo  {journal} {Reviews of Modern
  Physics}\ }\textbf {\bibinfo {volume} {87}},\ \bibinfo {pages} {483}
  (\bibinfo {year} {2015})}\BibitemShut {NoStop}%
\bibitem [{\citenamefont {Volpe}\ and\ \citenamefont
  {Volpe}(2017)}]{volpe2017topography}%
  \BibitemOpen
  \bibfield  {author} {\bibinfo {author} {\bibfnamefont {G.}~\bibnamefont
  {Volpe}}\ and\ \bibinfo {author} {\bibfnamefont {G.}~\bibnamefont {Volpe}},\
  }\bibfield  {title} {\enquote {\bibinfo {title} {The topography of the
  environment alters the optimal search strategy for active particles},}\
  }\href@noop {} {\bibfield  {journal} {\bibinfo  {journal} {Proceedings of the
  National Academy of Sciences}\ }\textbf {\bibinfo {volume} {114}},\ \bibinfo
  {pages} {11350--11355} (\bibinfo {year} {2017})}\BibitemShut {NoStop}%
\bibitem [{\citenamefont {Lomholt}\ \emph {et~al.}(2008)\citenamefont
  {Lomholt}, \citenamefont {Tal}, \citenamefont {Metzler},\ and\ \citenamefont
  {Joseph}}]{lomholt2008levy}%
  \BibitemOpen
  \bibfield  {author} {\bibinfo {author} {\bibfnamefont {M.~A.}\ \bibnamefont
  {Lomholt}}, \bibinfo {author} {\bibfnamefont {K.}~\bibnamefont {Tal}},
  \bibinfo {author} {\bibfnamefont {R.}~\bibnamefont {Metzler}}, \ and\
  \bibinfo {author} {\bibfnamefont {K.}~\bibnamefont {Joseph}},\ }\bibfield
  {title} {\enquote {\bibinfo {title} {L{\'e}vy strategies in intermittent
  search processes are advantageous},}\ }\href@noop {} {\bibfield  {journal}
  {\bibinfo  {journal} {Proceedings of the National Academy of Sciences}\
  }\textbf {\bibinfo {volume} {105}},\ \bibinfo {pages} {11055--11059}
  (\bibinfo {year} {2008})}\BibitemShut {NoStop}%
\bibitem [{\citenamefont {Palyulin}, \citenamefont {Chechkin},\ and\
  \citenamefont {Metzler}(2014)}]{palyulin2014levy}%
  \BibitemOpen
  \bibfield  {author} {\bibinfo {author} {\bibfnamefont {V.~V.}\ \bibnamefont
  {Palyulin}}, \bibinfo {author} {\bibfnamefont {A.~V.}\ \bibnamefont
  {Chechkin}}, \ and\ \bibinfo {author} {\bibfnamefont {R.}~\bibnamefont
  {Metzler}},\ }\bibfield  {title} {\enquote {\bibinfo {title} {L{\'e}vy
  flights do not always optimize random blind search for sparse targets},}\
  }\href@noop {} {\bibfield  {journal} {\bibinfo  {journal} {Proceedings of the
  National Academy of Sciences}\ }\textbf {\bibinfo {volume} {111}},\ \bibinfo
  {pages} {2931--2936} (\bibinfo {year} {2014})}\BibitemShut {NoStop}%
\bibitem [{\citenamefont {Evans}\ and\ \citenamefont
  {Majumdar}(2011{\natexlab{a}})}]{evans2011diffusion}%
  \BibitemOpen
  \bibfield  {author} {\bibinfo {author} {\bibfnamefont {M.~R.}\ \bibnamefont
  {Evans}}\ and\ \bibinfo {author} {\bibfnamefont {S.~N.}\ \bibnamefont
  {Majumdar}},\ }\bibfield  {title} {\enquote {\bibinfo {title} {Diffusion with
  stochastic resetting},}\ }\href@noop {} {\bibfield  {journal} {\bibinfo
  {journal} {Physical Review Letters}\ }\textbf {\bibinfo {volume} {106}},\
  \bibinfo {pages} {160601} (\bibinfo {year} {2011}{\natexlab{a}})}\BibitemShut
  {NoStop}%
\bibitem [{\citenamefont {Evans}\ and\ \citenamefont
  {Majumdar}(2011{\natexlab{b}})}]{evans2011diffusion-opt}%
  \BibitemOpen
  \bibfield  {author} {\bibinfo {author} {\bibfnamefont {M.~R.}\ \bibnamefont
  {Evans}}\ and\ \bibinfo {author} {\bibfnamefont {S.~N.}\ \bibnamefont
  {Majumdar}},\ }\bibfield  {title} {\enquote {\bibinfo {title} {Diffusion with
  optimal resetting},}\ }\href@noop {} {\bibfield  {journal} {\bibinfo
  {journal} {Journal of Physics A: Mathematical and Theoretical}\ }\textbf
  {\bibinfo {volume} {44}},\ \bibinfo {pages} {435001} (\bibinfo {year}
  {2011}{\natexlab{b}})}\BibitemShut {NoStop}%
\bibitem [{\citenamefont {Evans}, \citenamefont {Majumdar},\ and\ \citenamefont
  {Schehr}(2020)}]{evans2020stochastic}%
  \BibitemOpen
  \bibfield  {author} {\bibinfo {author} {\bibfnamefont {M.~R.}\ \bibnamefont
  {Evans}}, \bibinfo {author} {\bibfnamefont {S.~N.}\ \bibnamefont {Majumdar}},
  \ and\ \bibinfo {author} {\bibfnamefont {G.}~\bibnamefont {Schehr}},\
  }\bibfield  {title} {\enquote {\bibinfo {title} {Stochastic resetting and
  applications},}\ }\href@noop {} {\bibfield  {journal} {\bibinfo  {journal}
  {Journal of Physics A: Mathematical and Theoretical}\ }\textbf {\bibinfo
  {volume} {53}},\ \bibinfo {pages} {193001} (\bibinfo {year}
  {2020})}\BibitemShut {NoStop}%
\bibitem [{\citenamefont {Gupta}\ and\ \citenamefont
  {Jayannavar}(2022)}]{gupta2022stochastic}%
  \BibitemOpen
  \bibfield  {author} {\bibinfo {author} {\bibfnamefont {S.}~\bibnamefont
  {Gupta}}\ and\ \bibinfo {author} {\bibfnamefont {A.~M.}\ \bibnamefont
  {Jayannavar}},\ }\bibfield  {title} {\enquote {\bibinfo {title} {Stochastic
  resetting: A (very) brief review},}\ }\href@noop {} {\bibfield  {journal}
  {\bibinfo  {journal} {Frontiers in Physics}\ ,\ \bibinfo {pages} {130}}
  (\bibinfo {year} {2022})}\BibitemShut {NoStop}%
\bibitem [{\citenamefont {Kusmierz}\ \emph {et~al.}(2014)\citenamefont
  {Kusmierz}, \citenamefont {Majumdar}, \citenamefont {Sabhapandit},\ and\
  \citenamefont {Schehr}}]{kusmierz2014first}%
  \BibitemOpen
  \bibfield  {author} {\bibinfo {author} {\bibfnamefont {L.}~\bibnamefont
  {Kusmierz}}, \bibinfo {author} {\bibfnamefont {S.~N.}\ \bibnamefont
  {Majumdar}}, \bibinfo {author} {\bibfnamefont {S.}~\bibnamefont
  {Sabhapandit}}, \ and\ \bibinfo {author} {\bibfnamefont {G.}~\bibnamefont
  {Schehr}},\ }\bibfield  {title} {\enquote {\bibinfo {title} {First order
  transition for the optimal search time of l{\'e}vy flights with resetting},}\
  }\href@noop {} {\bibfield  {journal} {\bibinfo  {journal} {Physical Review
  Letters}\ }\textbf {\bibinfo {volume} {113}},\ \bibinfo {pages} {220602}
  (\bibinfo {year} {2014})}\BibitemShut {NoStop}%
\bibitem [{\citenamefont {Reuveni}(2016)}]{reuveni2016optimal}%
  \BibitemOpen
  \bibfield  {author} {\bibinfo {author} {\bibfnamefont {S.}~\bibnamefont
  {Reuveni}},\ }\bibfield  {title} {\enquote {\bibinfo {title} {Optimal
  stochastic restart renders fluctuations in first passage times universal},}\
  }\href@noop {} {\bibfield  {journal} {\bibinfo  {journal} {Physical Review
  Letters}\ }\textbf {\bibinfo {volume} {116}},\ \bibinfo {pages} {170601}
  (\bibinfo {year} {2016})}\BibitemShut {NoStop}%
\bibitem [{\citenamefont {Pal}\ and\ \citenamefont
  {Reuveni}(2017)}]{pal2017first}%
  \BibitemOpen
  \bibfield  {author} {\bibinfo {author} {\bibfnamefont {A.}~\bibnamefont
  {Pal}}\ and\ \bibinfo {author} {\bibfnamefont {S.}~\bibnamefont {Reuveni}},\
  }\bibfield  {title} {\enquote {\bibinfo {title} {First passage under
  restart},}\ }\href@noop {} {\bibfield  {journal} {\bibinfo  {journal}
  {Physical Review Letters}\ }\textbf {\bibinfo {volume} {118}},\ \bibinfo
  {pages} {030603} (\bibinfo {year} {2017})}\BibitemShut {NoStop}%
\bibitem [{\citenamefont {Pal}, \citenamefont {Eliazar},\ and\ \citenamefont
  {Reuveni}(2019)}]{pal2019first}%
  \BibitemOpen
  \bibfield  {author} {\bibinfo {author} {\bibfnamefont {A.}~\bibnamefont
  {Pal}}, \bibinfo {author} {\bibfnamefont {I.}~\bibnamefont {Eliazar}}, \ and\
  \bibinfo {author} {\bibfnamefont {S.}~\bibnamefont {Reuveni}},\ }\bibfield
  {title} {\enquote {\bibinfo {title} {First passage under restart with
  branching},}\ }\href@noop {} {\bibfield  {journal} {\bibinfo  {journal}
  {Physical Review Letters}\ }\textbf {\bibinfo {volume} {122}},\ \bibinfo
  {pages} {020602} (\bibinfo {year} {2019})}\BibitemShut {NoStop}%
\bibitem [{\citenamefont {De~Bruyne}, \citenamefont {Randon-Furling},\ and\
  \citenamefont {Redner}(2020)}]{de2020optimization}%
  \BibitemOpen
  \bibfield  {author} {\bibinfo {author} {\bibfnamefont {B.}~\bibnamefont
  {De~Bruyne}}, \bibinfo {author} {\bibfnamefont {J.}~\bibnamefont
  {Randon-Furling}}, \ and\ \bibinfo {author} {\bibfnamefont {S.}~\bibnamefont
  {Redner}},\ }\bibfield  {title} {\enquote {\bibinfo {title} {Optimization in
  first-passage resetting},}\ }\href@noop {} {\bibfield  {journal} {\bibinfo
  {journal} {Physical Review Letters}\ }\textbf {\bibinfo {volume} {125}},\
  \bibinfo {pages} {050602} (\bibinfo {year} {2020})}\BibitemShut {NoStop}%
\bibitem [{\citenamefont {Pal}, \citenamefont {Kundu},\ and\ \citenamefont
  {Evans}(2016)}]{pal2016diffusion-K}%
  \BibitemOpen
  \bibfield  {author} {\bibinfo {author} {\bibfnamefont {A.}~\bibnamefont
  {Pal}}, \bibinfo {author} {\bibfnamefont {A.}~\bibnamefont {Kundu}}, \ and\
  \bibinfo {author} {\bibfnamefont {M.~R.}\ \bibnamefont {Evans}},\ }\bibfield
  {title} {\enquote {\bibinfo {title} {Diffusion under time-dependent
  resetting},}\ }\href@noop {} {\bibfield  {journal} {\bibinfo  {journal}
  {Journal of Physics A: Mathematical and Theoretical}\ }\textbf {\bibinfo
  {volume} {49}},\ \bibinfo {pages} {225001} (\bibinfo {year}
  {2016})}\BibitemShut {NoStop}%
\bibitem [{\citenamefont {Chechkin}\ and\ \citenamefont
  {Sokolov}(2018)}]{chechkin2018random}%
  \BibitemOpen
  \bibfield  {author} {\bibinfo {author} {\bibfnamefont {A.}~\bibnamefont
  {Chechkin}}\ and\ \bibinfo {author} {\bibfnamefont {I.}~\bibnamefont
  {Sokolov}},\ }\bibfield  {title} {\enquote {\bibinfo {title} {Random search
  with resetting: a unified renewal approach},}\ }\href@noop {} {\bibfield
  {journal} {\bibinfo  {journal} {Physical Review Letters}\ }\textbf {\bibinfo
  {volume} {121}},\ \bibinfo {pages} {050601} (\bibinfo {year}
  {2018})}\BibitemShut {NoStop}%
\bibitem [{\citenamefont {Pal}\ and\ \citenamefont
  {Prasad}(2019{\natexlab{a}})}]{pal2019first-V}%
  \BibitemOpen
  \bibfield  {author} {\bibinfo {author} {\bibfnamefont {A.}~\bibnamefont
  {Pal}}\ and\ \bibinfo {author} {\bibfnamefont {V.~V.}\ \bibnamefont
  {Prasad}},\ }\bibfield  {title} {\enquote {\bibinfo {title} {First passage
  under stochastic resetting in an interval},}\ }\href@noop {} {\bibfield
  {journal} {\bibinfo  {journal} {Physical Review E}\ }\textbf {\bibinfo
  {volume} {99}},\ \bibinfo {pages} {032123} (\bibinfo {year}
  {2019}{\natexlab{a}})}\BibitemShut {NoStop}%
\bibitem [{\citenamefont {Bhat}, \citenamefont {De~Bacco},\ and\ \citenamefont
  {Redner}(2016)}]{bhat2016stochastic}%
  \BibitemOpen
  \bibfield  {author} {\bibinfo {author} {\bibfnamefont {U.}~\bibnamefont
  {Bhat}}, \bibinfo {author} {\bibfnamefont {C.}~\bibnamefont {De~Bacco}}, \
  and\ \bibinfo {author} {\bibfnamefont {S.}~\bibnamefont {Redner}},\
  }\bibfield  {title} {\enquote {\bibinfo {title} {Stochastic search with
  poisson and deterministic resetting},}\ }\href@noop {} {\bibfield  {journal}
  {\bibinfo  {journal} {Journal of Statistical Mechanics: Theory and
  Experiment}\ }\textbf {\bibinfo {volume} {2016}},\ \bibinfo {pages} {083401}
  (\bibinfo {year} {2016})}\BibitemShut {NoStop}%
\bibitem [{\citenamefont {Pal}\ and\ \citenamefont
  {Prasad}(2019{\natexlab{b}})}]{pal2019landau}%
  \BibitemOpen
  \bibfield  {author} {\bibinfo {author} {\bibfnamefont {A.}~\bibnamefont
  {Pal}}\ and\ \bibinfo {author} {\bibfnamefont {V.~V.}\ \bibnamefont
  {Prasad}},\ }\bibfield  {title} {\enquote {\bibinfo {title} {Landau-like
  expansion for phase transitions in stochastic resetting},}\ }\href@noop {}
  {\bibfield  {journal} {\bibinfo  {journal} {Physical Review Research}\
  }\textbf {\bibinfo {volume} {1}},\ \bibinfo {pages} {032001} (\bibinfo {year}
  {2019}{\natexlab{b}})}\BibitemShut {NoStop}%
\bibitem [{\citenamefont {Bressloff}(2020)}]{bressloff2020directed}%
  \BibitemOpen
  \bibfield  {author} {\bibinfo {author} {\bibfnamefont {P.~C.}\ \bibnamefont
  {Bressloff}},\ }\bibfield  {title} {\enquote {\bibinfo {title} {Directed
  intermittent search with stochastic resetting},}\ }\href@noop {} {\bibfield
  {journal} {\bibinfo  {journal} {Journal of Physics A: Mathematical and
  Theoretical}\ }\textbf {\bibinfo {volume} {53}},\ \bibinfo {pages} {105001}
  (\bibinfo {year} {2020})}\BibitemShut {NoStop}%
\bibitem [{\citenamefont {Tal-Friedman}\ \emph {et~al.}(2020)\citenamefont
  {Tal-Friedman}, \citenamefont {Pal}, \citenamefont {Sekhon}, \citenamefont
  {Reuveni},\ and\ \citenamefont {Roichman}}]{tal2020experimental}%
  \BibitemOpen
  \bibfield  {author} {\bibinfo {author} {\bibfnamefont {O.}~\bibnamefont
  {Tal-Friedman}}, \bibinfo {author} {\bibfnamefont {A.}~\bibnamefont {Pal}},
  \bibinfo {author} {\bibfnamefont {A.}~\bibnamefont {Sekhon}}, \bibinfo
  {author} {\bibfnamefont {S.}~\bibnamefont {Reuveni}}, \ and\ \bibinfo
  {author} {\bibfnamefont {Y.}~\bibnamefont {Roichman}},\ }\bibfield  {title}
  {\enquote {\bibinfo {title} {Experimental realization of diffusion with
  stochastic resetting},}\ }\href@noop {} {\bibfield  {journal} {\bibinfo
  {journal} {The Journal of Physical Chemistry Letters}\ }\textbf {\bibinfo
  {volume} {11}},\ \bibinfo {pages} {7350--7355} (\bibinfo {year}
  {2020})}\BibitemShut {NoStop}%
\bibitem [{\citenamefont {Montanari}\ and\ \citenamefont
  {Zecchina}(2002)}]{montanari2002optimizing}%
  \BibitemOpen
  \bibfield  {author} {\bibinfo {author} {\bibfnamefont {A.}~\bibnamefont
  {Montanari}}\ and\ \bibinfo {author} {\bibfnamefont {R.}~\bibnamefont
  {Zecchina}},\ }\bibfield  {title} {\enquote {\bibinfo {title} {Optimizing
  searches via rare events},}\ }\href@noop {} {\bibfield  {journal} {\bibinfo
  {journal} {Physical Review Letters}\ }\textbf {\bibinfo {volume} {88}},\
  \bibinfo {pages} {178701} (\bibinfo {year} {2002})}\BibitemShut {NoStop}%
\bibitem [{\citenamefont {Luby}, \citenamefont {Sinclair},\ and\ \citenamefont
  {Zuckerman}(1993)}]{luby1993optimal}%
  \BibitemOpen
  \bibfield  {author} {\bibinfo {author} {\bibfnamefont {M.}~\bibnamefont
  {Luby}}, \bibinfo {author} {\bibfnamefont {A.}~\bibnamefont {Sinclair}}, \
  and\ \bibinfo {author} {\bibfnamefont {D.}~\bibnamefont {Zuckerman}},\
  }\bibfield  {title} {\enquote {\bibinfo {title} {Optimal speedup of las vegas
  algorithms},}\ }\href@noop {} {\bibfield  {journal} {\bibinfo  {journal}
  {Information Processing Letters}\ }\textbf {\bibinfo {volume} {47}},\
  \bibinfo {pages} {173--180} (\bibinfo {year} {1993})}\BibitemShut {NoStop}%
\bibitem [{\citenamefont {Reuveni}, \citenamefont {Urbakh},\ and\ \citenamefont
  {Klafter}(2014)}]{reuveni2014role}%
  \BibitemOpen
  \bibfield  {author} {\bibinfo {author} {\bibfnamefont {S.}~\bibnamefont
  {Reuveni}}, \bibinfo {author} {\bibfnamefont {M.}~\bibnamefont {Urbakh}}, \
  and\ \bibinfo {author} {\bibfnamefont {J.}~\bibnamefont {Klafter}},\
  }\bibfield  {title} {\enquote {\bibinfo {title} {Role of substrate unbinding
  in michaelis--menten enzymatic reactions},}\ }\href@noop {} {\bibfield
  {journal} {\bibinfo  {journal} {Proceedings of the National Academy of
  Sciences}\ }\textbf {\bibinfo {volume} {111}},\ \bibinfo {pages} {4391--4396}
  (\bibinfo {year} {2014})}\BibitemShut {NoStop}%
\bibitem [{\citenamefont {Ray}(2020)}]{ray2020space}%
  \BibitemOpen
  \bibfield  {author} {\bibinfo {author} {\bibfnamefont {S.}~\bibnamefont
  {Ray}},\ }\bibfield  {title} {\enquote {\bibinfo {title} {Space-dependent
  diffusion with stochastic resetting: A first-passage study},}\ }\href@noop {}
  {\bibfield  {journal} {\bibinfo  {journal} {The Journal of Chemical Physics}\
  }\textbf {\bibinfo {volume} {153}},\ \bibinfo {pages} {234904} (\bibinfo
  {year} {2020})}\BibitemShut {NoStop}%
\bibitem [{\citenamefont {Ray}\ \emph {et~al.}(2021)\citenamefont {Ray},
  \citenamefont {Pal}, \citenamefont {Ghosh}, \citenamefont {Dana},\ and\
  \citenamefont {Hens}}]{ray2021mitigating}%
  \BibitemOpen
  \bibfield  {author} {\bibinfo {author} {\bibfnamefont {A.}~\bibnamefont
  {Ray}}, \bibinfo {author} {\bibfnamefont {A.}~\bibnamefont {Pal}}, \bibinfo
  {author} {\bibfnamefont {D.}~\bibnamefont {Ghosh}}, \bibinfo {author}
  {\bibfnamefont {S.~K.}\ \bibnamefont {Dana}}, \ and\ \bibinfo {author}
  {\bibfnamefont {C.}~\bibnamefont {Hens}},\ }\bibfield  {title} {\enquote
  {\bibinfo {title} {Mitigating long transient time in deterministic systems by
  resetting},}\ }\href@noop {} {\bibfield  {journal} {\bibinfo  {journal}
  {Chaos: An Interdisciplinary Journal of Nonlinear Science}\ }\textbf
  {\bibinfo {volume} {31}},\ \bibinfo {pages} {011103} (\bibinfo {year}
  {2021})}\BibitemShut {NoStop}%
\bibitem [{\citenamefont {Rold{\'a}n}\ \emph {et~al.}(2016)\citenamefont
  {Rold{\'a}n}, \citenamefont {Lisica}, \citenamefont {S{\'a}nchez-Taltavull},\
  and\ \citenamefont {Grill}}]{roldan2016stochastic}%
  \BibitemOpen
  \bibfield  {author} {\bibinfo {author} {\bibfnamefont {{\'E}.}~\bibnamefont
  {Rold{\'a}n}}, \bibinfo {author} {\bibfnamefont {A.}~\bibnamefont {Lisica}},
  \bibinfo {author} {\bibfnamefont {D.}~\bibnamefont {S{\'a}nchez-Taltavull}},
  \ and\ \bibinfo {author} {\bibfnamefont {S.~W.}\ \bibnamefont {Grill}},\
  }\bibfield  {title} {\enquote {\bibinfo {title} {Stochastic resetting in
  backtrack recovery by rna polymerases},}\ }\href@noop {} {\bibfield
  {journal} {\bibinfo  {journal} {Physical Review E}\ }\textbf {\bibinfo
  {volume} {93}},\ \bibinfo {pages} {062411} (\bibinfo {year}
  {2016})}\BibitemShut {NoStop}%
\bibitem [{\citenamefont {Bonomo}, \citenamefont {Pal},\ and\ \citenamefont
  {Reuveni}(2022)}]{bonomo2021mitigating}%
  \BibitemOpen
  \bibfield  {author} {\bibinfo {author} {\bibfnamefont {O.~L.}\ \bibnamefont
  {Bonomo}}, \bibinfo {author} {\bibfnamefont {A.}~\bibnamefont {Pal}}, \ and\
  \bibinfo {author} {\bibfnamefont {S.}~\bibnamefont {Reuveni}},\ }\bibfield
  {title} {\enquote {\bibinfo {title} {Mitigating long queues and waiting times
  with service resetting},}\ }\href@noop {} {\bibfield  {journal} {\bibinfo
  {journal} {PNAS Nexus}\ }\textbf {\bibinfo {volume} {1}},\ \bibinfo {pages}
  {pgac070} (\bibinfo {year} {2022})}\BibitemShut {NoStop}%
\bibitem [{\citenamefont {Pal}, \citenamefont {Ku{\'s}mierz},\ and\
  \citenamefont {Reuveni}(2020)}]{pal2020search}%
  \BibitemOpen
  \bibfield  {author} {\bibinfo {author} {\bibfnamefont {A.}~\bibnamefont
  {Pal}}, \bibinfo {author} {\bibfnamefont {{\L}.}~\bibnamefont
  {Ku{\'s}mierz}}, \ and\ \bibinfo {author} {\bibfnamefont {S.}~\bibnamefont
  {Reuveni}},\ }\bibfield  {title} {\enquote {\bibinfo {title} {Search with
  home returns provides advantage under high uncertainty},}\ }\href@noop {}
  {\bibfield  {journal} {\bibinfo  {journal} {Physical Review Research}\
  }\textbf {\bibinfo {volume} {2}},\ \bibinfo {pages} {043174} (\bibinfo {year}
  {2020})}\BibitemShut {NoStop}%
\bibitem [{\citenamefont {Campos}\ and\ \citenamefont
  {M{\'e}ndez}(2015)}]{campos2015phase}%
  \BibitemOpen
  \bibfield  {author} {\bibinfo {author} {\bibfnamefont {D.}~\bibnamefont
  {Campos}}\ and\ \bibinfo {author} {\bibfnamefont {V.}~\bibnamefont
  {M{\'e}ndez}},\ }\bibfield  {title} {\enquote {\bibinfo {title} {Phase
  transitions in optimal search times: How random walkers should combine
  resetting and flight scales},}\ }\href@noop {} {\bibfield  {journal}
  {\bibinfo  {journal} {Physical Review E}\ }\textbf {\bibinfo {volume} {92}},\
  \bibinfo {pages} {062115} (\bibinfo {year} {2015})}\BibitemShut {NoStop}%
\bibitem [{\citenamefont {Ku{\'s}mierz}\ and\ \citenamefont
  {Gudowska-Nowak}(2015)}]{kusmierz2015optimal}%
  \BibitemOpen
  \bibfield  {author} {\bibinfo {author} {\bibfnamefont {{\L}.}~\bibnamefont
  {Ku{\'s}mierz}}\ and\ \bibinfo {author} {\bibfnamefont {E.}~\bibnamefont
  {Gudowska-Nowak}},\ }\bibfield  {title} {\enquote {\bibinfo {title} {Optimal
  first-arrival times in l{\'e}vy flights with resetting},}\ }\href@noop {}
  {\bibfield  {journal} {\bibinfo  {journal} {Physical Review E}\ }\textbf
  {\bibinfo {volume} {92}},\ \bibinfo {pages} {052127} (\bibinfo {year}
  {2015})}\BibitemShut {NoStop}%
\bibitem [{\citenamefont {Singh}, \citenamefont {Metzler},\ and\ \citenamefont
  {Sandev}(2020)}]{singh2020resetting}%
  \BibitemOpen
  \bibfield  {author} {\bibinfo {author} {\bibfnamefont {R.}~\bibnamefont
  {Singh}}, \bibinfo {author} {\bibfnamefont {R.}~\bibnamefont {Metzler}}, \
  and\ \bibinfo {author} {\bibfnamefont {T.}~\bibnamefont {Sandev}},\
  }\bibfield  {title} {\enquote {\bibinfo {title} {Resetting dynamics in a
  confining potential},}\ }\href@noop {} {\bibfield  {journal} {\bibinfo
  {journal} {Journal of Physics A: Mathematical and Theoretical}\ }\textbf
  {\bibinfo {volume} {53}},\ \bibinfo {pages} {505003} (\bibinfo {year}
  {2020})}\BibitemShut {NoStop}%
\bibitem [{\citenamefont {Pal}(2015)}]{pal2015diffusion}%
  \BibitemOpen
  \bibfield  {author} {\bibinfo {author} {\bibfnamefont {A.}~\bibnamefont
  {Pal}},\ }\bibfield  {title} {\enquote {\bibinfo {title} {Diffusion in a
  potential landscape with stochastic resetting},}\ }\href@noop {} {\bibfield
  {journal} {\bibinfo  {journal} {Physical Review E}\ }\textbf {\bibinfo
  {volume} {91}},\ \bibinfo {pages} {012113} (\bibinfo {year}
  {2015})}\BibitemShut {NoStop}%
\bibitem [{\citenamefont {Ray}, \citenamefont {Mondal},\ and\ \citenamefont
  {Reuveni}(2019)}]{ray2019peclet}%
  \BibitemOpen
  \bibfield  {author} {\bibinfo {author} {\bibfnamefont {S.}~\bibnamefont
  {Ray}}, \bibinfo {author} {\bibfnamefont {D.}~\bibnamefont {Mondal}}, \ and\
  \bibinfo {author} {\bibfnamefont {S.}~\bibnamefont {Reuveni}},\ }\bibfield
  {title} {\enquote {\bibinfo {title} {P{\'e}clet number governs transition to
  acceleratory restart in drift-diffusion},}\ }\href@noop {} {\bibfield
  {journal} {\bibinfo  {journal} {Journal of Physics A: Mathematical and
  Theoretical}\ }\textbf {\bibinfo {volume} {52}},\ \bibinfo {pages} {255002}
  (\bibinfo {year} {2019})}\BibitemShut {NoStop}%
\bibitem [{\citenamefont {Huang}\ and\ \citenamefont
  {Chen}(2021)}]{huang2021random}%
  \BibitemOpen
  \bibfield  {author} {\bibinfo {author} {\bibfnamefont {F.}~\bibnamefont
  {Huang}}\ and\ \bibinfo {author} {\bibfnamefont {H.}~\bibnamefont {Chen}},\
  }\bibfield  {title} {\enquote {\bibinfo {title} {Random walks on complex
  networks with first-passage resetting},}\ }\href@noop {} {\bibfield
  {journal} {\bibinfo  {journal} {Physical Review E}\ }\textbf {\bibinfo
  {volume} {103}},\ \bibinfo {pages} {062132} (\bibinfo {year}
  {2021})}\BibitemShut {NoStop}%
\bibitem [{\citenamefont {Gonz{\'a}lez}, \citenamefont {Riascos},\ and\
  \citenamefont {Boyer}(2021)}]{gonzalez2021diffusive}%
  \BibitemOpen
  \bibfield  {author} {\bibinfo {author} {\bibfnamefont {F.~H.}\ \bibnamefont
  {Gonz{\'a}lez}}, \bibinfo {author} {\bibfnamefont {A.~P.}\ \bibnamefont
  {Riascos}}, \ and\ \bibinfo {author} {\bibfnamefont {D.}~\bibnamefont
  {Boyer}},\ }\bibfield  {title} {\enquote {\bibinfo {title} {Diffusive
  transport on networks with stochastic resetting to multiple nodes},}\
  }\href@noop {} {\bibfield  {journal} {\bibinfo  {journal} {Physical Review
  E}\ }\textbf {\bibinfo {volume} {103}},\ \bibinfo {pages} {062126} (\bibinfo
  {year} {2021})}\BibitemShut {NoStop}%
\bibitem [{\citenamefont {Bonomo}\ and\ \citenamefont
  {Pal}(2021)}]{bonomo2021first}%
  \BibitemOpen
  \bibfield  {author} {\bibinfo {author} {\bibfnamefont {O.~L.}\ \bibnamefont
  {Bonomo}}\ and\ \bibinfo {author} {\bibfnamefont {A.}~\bibnamefont {Pal}},\
  }\bibfield  {title} {\enquote {\bibinfo {title} {First passage under restart
  for discrete space and time: Application to one-dimensional confined lattice
  random walks},}\ }\href@noop {} {\bibfield  {journal} {\bibinfo  {journal}
  {Physical Review E}\ }\textbf {\bibinfo {volume} {103}},\ \bibinfo {pages}
  {052129} (\bibinfo {year} {2021})}\BibitemShut {NoStop}%
\bibitem [{\citenamefont {Pal}, \citenamefont {Kostinski},\ and\ \citenamefont
  {Reuveni}(2022)}]{pal2022inspection}%
  \BibitemOpen
  \bibfield  {author} {\bibinfo {author} {\bibfnamefont {A.}~\bibnamefont
  {Pal}}, \bibinfo {author} {\bibfnamefont {S.}~\bibnamefont {Kostinski}}, \
  and\ \bibinfo {author} {\bibfnamefont {S.}~\bibnamefont {Reuveni}},\
  }\bibfield  {title} {\enquote {\bibinfo {title} {The inspection paradox in
  stochastic resetting},}\ }\href@noop {} {\bibfield  {journal} {\bibinfo
  {journal} {Journal of Physics A: Mathematical and Theoretical}\ }\textbf
  {\bibinfo {volume} {55}},\ \bibinfo {pages} {021001} (\bibinfo {year}
  {2022})}\BibitemShut {NoStop}%
\bibitem [{\citenamefont {Volpe}, \citenamefont {Gigan},\ and\ \citenamefont
  {Volpe}(2014)}]{volpe2014simulation}%
  \BibitemOpen
  \bibfield  {author} {\bibinfo {author} {\bibfnamefont {G.}~\bibnamefont
  {Volpe}}, \bibinfo {author} {\bibfnamefont {S.}~\bibnamefont {Gigan}}, \ and\
  \bibinfo {author} {\bibfnamefont {G.}~\bibnamefont {Volpe}},\ }\bibfield
  {title} {\enquote {\bibinfo {title} {Simulation of the active brownian motion
  of a microswimmer},}\ }\href@noop {} {\bibfield  {journal} {\bibinfo
  {journal} {American Journal of Physics}\ }\textbf {\bibinfo {volume} {82}},\
  \bibinfo {pages} {659--664} (\bibinfo {year} {2014})}\BibitemShut {NoStop}%
\bibitem [{\citenamefont {Basu}\ \emph {et~al.}(2018)\citenamefont {Basu},
  \citenamefont {Majumdar}, \citenamefont {Rosso},\ and\ \citenamefont
  {Schehr}}]{basu2018active}%
  \BibitemOpen
  \bibfield  {author} {\bibinfo {author} {\bibfnamefont {U.}~\bibnamefont
  {Basu}}, \bibinfo {author} {\bibfnamefont {S.~N.}\ \bibnamefont {Majumdar}},
  \bibinfo {author} {\bibfnamefont {A.}~\bibnamefont {Rosso}}, \ and\ \bibinfo
  {author} {\bibfnamefont {G.}~\bibnamefont {Schehr}},\ }\bibfield  {title}
  {\enquote {\bibinfo {title} {Active brownian motion in two dimensions},}\
  }\href@noop {} {\bibfield  {journal} {\bibinfo  {journal} {Physical Review
  E}\ }\textbf {\bibinfo {volume} {98}},\ \bibinfo {pages} {062121} (\bibinfo
  {year} {2018})}\BibitemShut {NoStop}%
\bibitem [{\citenamefont {Topaz}\ and\ \citenamefont
  {Bertozzi}(2004)}]{topaz2004swarming}%
  \BibitemOpen
  \bibfield  {author} {\bibinfo {author} {\bibfnamefont {C.~M.}\ \bibnamefont
  {Topaz}}\ and\ \bibinfo {author} {\bibfnamefont {A.~L.}\ \bibnamefont
  {Bertozzi}},\ }\bibfield  {title} {\enquote {\bibinfo {title} {Swarming
  patterns in a two-dimensional kinematic model for biological groups},}\
  }\href@noop {} {\bibfield  {journal} {\bibinfo  {journal} {SIAM Journal on
  Applied Mathematics}\ }\textbf {\bibinfo {volume} {65}},\ \bibinfo {pages}
  {152--174} (\bibinfo {year} {2004})}\BibitemShut {NoStop}%
\bibitem [{\citenamefont {Strefler}, \citenamefont {Erdmann},\ and\
  \citenamefont {Schimansky-Geier}(2008)}]{strefler2008swarming}%
  \BibitemOpen
  \bibfield  {author} {\bibinfo {author} {\bibfnamefont {J.}~\bibnamefont
  {Strefler}}, \bibinfo {author} {\bibfnamefont {U.}~\bibnamefont {Erdmann}}, \
  and\ \bibinfo {author} {\bibfnamefont {L.}~\bibnamefont {Schimansky-Geier}},\
  }\bibfield  {title} {\enquote {\bibinfo {title} {Swarming in three
  dimensions},}\ }\href@noop {} {\bibfield  {journal} {\bibinfo  {journal}
  {Physical Review E}\ }\textbf {\bibinfo {volume} {78}},\ \bibinfo {pages}
  {031927} (\bibinfo {year} {2008})}\BibitemShut {NoStop}%
\bibitem [{\citenamefont {Slowman}, \citenamefont {Evans},\ and\ \citenamefont
  {Blythe}(2016)}]{slowman2016jamming}%
  \BibitemOpen
  \bibfield  {author} {\bibinfo {author} {\bibfnamefont {A.}~\bibnamefont
  {Slowman}}, \bibinfo {author} {\bibfnamefont {M.}~\bibnamefont {Evans}}, \
  and\ \bibinfo {author} {\bibfnamefont {R.}~\bibnamefont {Blythe}},\
  }\bibfield  {title} {\enquote {\bibinfo {title} {Jamming and attraction of
  interacting run-and-tumble random walkers},}\ }\href@noop {} {\bibfield
  {journal} {\bibinfo  {journal} {Physical Review Letters}\ }\textbf {\bibinfo
  {volume} {116}},\ \bibinfo {pages} {218101} (\bibinfo {year}
  {2016})}\BibitemShut {NoStop}%
\bibitem [{\citenamefont {Palacci}\ \emph {et~al.}(2013)\citenamefont
  {Palacci}, \citenamefont {Sacanna}, \citenamefont {Steinberg}, \citenamefont
  {Pine},\ and\ \citenamefont {Chaikin}}]{palacci2013living}%
  \BibitemOpen
  \bibfield  {author} {\bibinfo {author} {\bibfnamefont {J.}~\bibnamefont
  {Palacci}}, \bibinfo {author} {\bibfnamefont {S.}~\bibnamefont {Sacanna}},
  \bibinfo {author} {\bibfnamefont {A.~P.}\ \bibnamefont {Steinberg}}, \bibinfo
  {author} {\bibfnamefont {D.~J.}\ \bibnamefont {Pine}}, \ and\ \bibinfo
  {author} {\bibfnamefont {P.~M.}\ \bibnamefont {Chaikin}},\ }\bibfield
  {title} {\enquote {\bibinfo {title} {Living crystals of light-activated
  colloidal surfers},}\ }\href@noop {} {\bibfield  {journal} {\bibinfo
  {journal} {Science}\ }\textbf {\bibinfo {volume} {339}},\ \bibinfo {pages}
  {936--940} (\bibinfo {year} {2013})}\BibitemShut {NoStop}%
\bibitem [{\citenamefont {Friedrich}\ and\ \citenamefont
  {J{\"u}licher}(2008)}]{friedrich2008stochastic}%
  \BibitemOpen
  \bibfield  {author} {\bibinfo {author} {\bibfnamefont {B.}~\bibnamefont
  {Friedrich}}\ and\ \bibinfo {author} {\bibfnamefont {F.}~\bibnamefont
  {J{\"u}licher}},\ }\bibfield  {title} {\enquote {\bibinfo {title} {The
  stochastic dance of circling sperm cells: sperm chemotaxis in the plane},}\
  }\href@noop {} {\bibfield  {journal} {\bibinfo  {journal} {New Journal of
  Physics}\ }\textbf {\bibinfo {volume} {10}},\ \bibinfo {pages} {123025}
  (\bibinfo {year} {2008})}\BibitemShut {NoStop}%
\bibitem [{\citenamefont {Redner}(2001)}]{redner2001guide}%
  \BibitemOpen
  \bibfield  {author} {\bibinfo {author} {\bibfnamefont {S.}~\bibnamefont
  {Redner}},\ }\href@noop {} {\emph {\bibinfo {title} {A guide to first-passage
  processes}}}\ (\bibinfo  {publisher} {Cambridge University Press},\ \bibinfo
  {year} {2001})\BibitemShut {NoStop}%
\bibitem [{\citenamefont {Evans}\ and\ \citenamefont
  {Majumdar}(2018)}]{evans2018run}%
  \BibitemOpen
  \bibfield  {author} {\bibinfo {author} {\bibfnamefont {M.~R.}\ \bibnamefont
  {Evans}}\ and\ \bibinfo {author} {\bibfnamefont {S.~N.}\ \bibnamefont
  {Majumdar}},\ }\bibfield  {title} {\enquote {\bibinfo {title} {Run and tumble
  particle under resetting: a renewal approach},}\ }\href@noop {} {\bibfield
  {journal} {\bibinfo  {journal} {Journal of Physics A: Mathematical and
  Theoretical}\ }\textbf {\bibinfo {volume} {51}},\ \bibinfo {pages} {475003}
  (\bibinfo {year} {2018})}\BibitemShut {NoStop}%
\bibitem [{\citenamefont {Kumar}, \citenamefont {Sadekar},\ and\ \citenamefont
  {Basu}(2020)}]{kumar2020active}%
  \BibitemOpen
  \bibfield  {author} {\bibinfo {author} {\bibfnamefont {V.}~\bibnamefont
  {Kumar}}, \bibinfo {author} {\bibfnamefont {O.}~\bibnamefont {Sadekar}}, \
  and\ \bibinfo {author} {\bibfnamefont {U.}~\bibnamefont {Basu}},\ }\bibfield
  {title} {\enquote {\bibinfo {title} {Active brownian motion in two dimensions
  under stochastic resetting},}\ }\href@noop {} {\bibfield  {journal} {\bibinfo
   {journal} {Physical Review E}\ }\textbf {\bibinfo {volume} {102}},\ \bibinfo
  {pages} {052129} (\bibinfo {year} {2020})}\BibitemShut {NoStop}%
\bibitem [{\citenamefont {Santra}, \citenamefont {Basu},\ and\ \citenamefont
  {Sabhapandit}(2020)}]{santra2020run}%
  \BibitemOpen
  \bibfield  {author} {\bibinfo {author} {\bibfnamefont {I.}~\bibnamefont
  {Santra}}, \bibinfo {author} {\bibfnamefont {U.}~\bibnamefont {Basu}}, \ and\
  \bibinfo {author} {\bibfnamefont {S.}~\bibnamefont {Sabhapandit}},\
  }\bibfield  {title} {\enquote {\bibinfo {title} {Run-and-tumble particles in
  two dimensions under stochastic resetting conditions},}\ }\href@noop {}
  {\bibfield  {journal} {\bibinfo  {journal} {Journal of Statistical Mechanics:
  Theory and Experiment}\ }\textbf {\bibinfo {volume} {2020}},\ \bibinfo
  {pages} {113206} (\bibinfo {year} {2020})}\BibitemShut {NoStop}%
\bibitem [{\citenamefont {Abdoli}\ and\ \citenamefont
  {Sharma}(2021)}]{abdoli2021stochastic}%
  \BibitemOpen
  \bibfield  {author} {\bibinfo {author} {\bibfnamefont {I.}~\bibnamefont
  {Abdoli}}\ and\ \bibinfo {author} {\bibfnamefont {A.}~\bibnamefont
  {Sharma}},\ }\bibfield  {title} {\enquote {\bibinfo {title} {Stochastic
  resetting of active brownian particles with lorentz force},}\ }\href@noop {}
  {\bibfield  {journal} {\bibinfo  {journal} {Soft Matter}\ }\textbf {\bibinfo
  {volume} {17}},\ \bibinfo {pages} {1307--1316} (\bibinfo {year}
  {2021})}\BibitemShut {NoStop}%
\bibitem [{\citenamefont {Besga}\ \emph {et~al.}(2020)\citenamefont {Besga},
  \citenamefont {Bovon}, \citenamefont {Petrosyan}, \citenamefont {Majumdar},\
  and\ \citenamefont {Ciliberto}}]{besga2020optimal}%
  \BibitemOpen
  \bibfield  {author} {\bibinfo {author} {\bibfnamefont {B.}~\bibnamefont
  {Besga}}, \bibinfo {author} {\bibfnamefont {A.}~\bibnamefont {Bovon}},
  \bibinfo {author} {\bibfnamefont {A.}~\bibnamefont {Petrosyan}}, \bibinfo
  {author} {\bibfnamefont {S.~N.}\ \bibnamefont {Majumdar}}, \ and\ \bibinfo
  {author} {\bibfnamefont {S.}~\bibnamefont {Ciliberto}},\ }\bibfield  {title}
  {\enquote {\bibinfo {title} {Optimal mean first-passage time for a brownian
  searcher subjected to resetting: experimental and theoretical results},}\
  }\href@noop {} {\bibfield  {journal} {\bibinfo  {journal} {Physical Review
  Research}\ }\textbf {\bibinfo {volume} {2}},\ \bibinfo {pages} {032029}
  (\bibinfo {year} {2020})}\BibitemShut {NoStop}%
\bibitem [{sar()}]{sar2022}%
  \BibitemOpen
  \href@noop {} {}\bibinfo {howpublished}
  {\url{https://github.com/gourab-sar/ABP-under-reset}}\BibitemShut {NoStop}%
\bibitem [{\citenamefont {van Roon}\ \emph {et~al.}(2022)\citenamefont {van
  Roon}, \citenamefont {Volpe}, \citenamefont {da~Gama},\ and\ \citenamefont
  {Ara{\'u}jo}}]{van2022role}%
  \BibitemOpen
  \bibfield  {author} {\bibinfo {author} {\bibfnamefont {D.~M.}\ \bibnamefont
  {van Roon}}, \bibinfo {author} {\bibfnamefont {G.}~\bibnamefont {Volpe}},
  \bibinfo {author} {\bibfnamefont {M.~M.~T.}\ \bibnamefont {da~Gama}}, \ and\
  \bibinfo {author} {\bibfnamefont {N.~A.}\ \bibnamefont {Ara{\'u}jo}},\
  }\bibfield  {title} {\enquote {\bibinfo {title} {The role of disorder in the
  motion of chiral swimmers in the presence of obstacles},}\ }\href@noop {}
  {\bibfield  {journal} {\bibinfo  {journal} {arXiv preprint arXiv:2205.13509}\
  } (\bibinfo {year} {2022})}\BibitemShut {NoStop}%
\end{thebibliography}%

%\end{thebibliography}

\end{document}